\documentclass[pdflatex,sn-nature]{sn-jnl}

\usepackage{graphicx}%
\usepackage{multirow}%
\usepackage{amsmath,amssymb,amsfonts}%
\usepackage{amsthm}%
\usepackage{mathrsfs}%
\usepackage[title]{appendix}%
\usepackage{xcolor}%
\usepackage{textcomp}%
\usepackage{manyfoot}%
\usepackage{booktabs}%
\usepackage{algorithm}%
\usepackage{algorithmicx}%
\usepackage{algpseudocode}%
\usepackage{listings}%
\usepackage{xspace}
\usepackage[T1]{fontenc}

\setcitestyle{super}
\makeatletter
\def\@cite#1#2{#1\if@tempswa , #2\fi}
\renewcommand\NAT@open{}  
\renewcommand\NAT@close{} 
\makeatother

\newcommand{\Rj}{\ensuremath{R_{\rm{Jup}}}\xspace}
\newcommand{\Mj}{\ensuremath{M_{\rm{Jup}}}\xspace}

\newcommand{\Lsun}{L_\odot}
\newcommand{\Teff}{\ensuremath{T_\mathrm{eff}}\xspace}

\newcommand{\lbollsun}{\ensuremath{\log(L_\mathrm{bol}/\Lsun)}}
\newcommand{\lbol}{\ensuremath{L_\mathrm{bol}}}

\newcommand{\logco}{\ensuremath{\mathrm{log(^{12}CO / ^{13}CO)}\xspace}}
\newcommand{\logcoo}{\ensuremath{\mathrm{log(C^{16}O / C^{18}O)}\xspace}}

\newcommand{\co}{\ensuremath{\mathrm{^{12}CO / ^{13}CO}\xspace}}
\newcommand{\coo}{\ensuremath{\mathrm{C^{16}O / C^{18}O}\xspace}}
\newcommand{\kms}{km~s$^{-1}$\xspace}

\theoremstyle{thmstyleone}%
\theoremstyle{thmstyletwo}%

\theoremstyle{thmstylethree}%

\begin{document}

\title[Article Title]{Jupiter-like uniform metal enrichment in a system of multiple giant exoplanets}

\author*[1]{\fnm{Jean-Baptiste} \sur{Ruffio}}\email{jruffio@ucsd.edu}
\equalcont{These authors contributed equally to this work.}

\author[2]{\fnm{Jerry W.} \sur{Xuan}}
\equalcont{These authors contributed equally to this work.}

\author[3]{\fnm{Yayaati} \sur{Chachan}}

\author[4]{\fnm{Aurora} \sur{Kesseli}}

\author[1]{\fnm{Eve J.} \sur{Lee}}

\author[4]{\fnm{Charles} \sur{Beichman}}
 
\author[5]{\fnm{Klaus} \sur{Hodapp}}

\author[6,7]{\fnm{William O.} \sur{Balmer}}

\author[1]{\fnm{Quinn} \sur{Konopacky}}

\author[7]{\fnm{Marshall D.} \sur{Perrin}}

\author[2,8]{\fnm{Dimitri} \sur{Mawet}}

\author[9]{\fnm{Heather A.} \sur{Knutson}}

\author[4]{\fnm{Geoffrey} \sur{Bryden}}

\author[4,10]{\fnm{Thomas P.} \sur{Greene}}

\author[11,12]{\fnm{Doug} \sur{Johnstone}}

\author[13]{\fnm{Jarron} \sur{Leisenring}}

\author[14]{\fnm{Michael} \sur{Meyer}}

\author[8]{\fnm{Marie} \sur{Ygouf}}

\affil*[1]{Department of Astronomy \& Astrophysics, University of California San Diego, La Jolla, CA, USA}

\affil[2]{Department of Astronomy, California Institute of Technology, Pasadena, CA, USA}

\affil[3]{Department of Astronomy and Astrophysics, University of California, Santa Cruz, Santa
Cruz, CA, USA}

\affil[4]{IPAC, Caltech, Pasadena, CA, USA}

\affil[5]{University of Hawaii, Institute for Astronomy, 640 N. A'ohoku Place, Hilo, HI 96720, USA}

\affil[6]{Department of Physics \& Astronomy, Johns Hopkins University, 3400 N. Charles Street, Baltimore, MD 21218, USA}

\affil[7]{Space Telescope Science Institute, 3700 San Martin Drive, Baltimore, MD 21218, USA}

\affil[8]{Jet Propulsion
Laboratory, California Institute of Technology, Pasadena, CA, USA}

\affil[9]{Division of
Geological and Planetary Sciences, California Institute of Technology, Pasadena, CA, USA}

\affil[10]{Space Science and Astrobiology Division, NASA's Ames Research Center, Moffett Field, CA 94035, USA}

\affil[11]{NRC Herzberg Astronomy and Astrophysics, 5071 West Saanich Road, Victoria, BC, V9E 2E7, Canada}

\affil[12]{Department of Physics and Astronomy, University of Victoria, 3800 Finnerty Road, Elliot Building, Victoria, BC, V8P 5C2, Canada}

\affil[13]{Department of Astronomy/Steward Observatory, University of Arizona, 933 N. Cherry Ave., Tucson, AZ 85721, USA}

\affil[14]{Department of Astronomy, University of Michigan, Ann Arbor, MI 48109, USA}

\abstract{
The accretion of icy and rocky solids during the formation of a gas giant planet is poorly constrained and challenging to model. Refractory species, like sulfur, are only present in solids in the protoplanetary disk where planets form. Measuring their abundance in planetary atmospheres is one of the most direct ways of constraining the extent and mechanism of solid accretion.
Using the unprecedented sensitivity of NASA's James Webb Space Telescope (JWST), we measure a detailed chemical make-up of three massive gas giants orbiting the star HR~8799 including direct detections of H$_2$O, CO, CH$_4$, CO$_2$, H$_2$S, $^{13}$CO, and C$^{18}$O. 
We find these planets are uniformly and highly enriched in heavy elements compared to the star irrespective of their volatile (carbon and oxygen) or refractory (sulfur) nature, which strongly suggests efficient accretion of solids during their formation.
This composition closely resembles that of Jupiter and Saturn and demonstrates that this enrichment also occurs in systems of multiple gas giant planets orbiting stars beyond the Solar System. This discovery hints at a shared origin for the heavy element enrichment of giant planets across a wider range of planet masses and orbital separations than previously anticipated.
}

\maketitle

HR~8799 is the first multi-planet system detected through direct imaging \cite{Marois2008science}.
This system features four gas-giant planets orbiting their star at a distance of 15-70 astronomical units and with individual planet masses between $5-10$ Jupiter masses ($\mathrm{M}_{\mathrm{Jup}}$) \cite{Zurlo2022}.
This iconic system has posed a fascinating challenge for planet formation theories. Indeed, the accretion timescale for classical core accretion increases as the cube of a planet's distance to its star \cite{Ida2004}, making the formation of distant gas giants similar to HR~8799 challenging. 
It also remains unclear how such massive planets could have formed so far away from their star given the vast amounts of solid materials needed in the disk to form their cores and enrich their atmospheres in heavy elements \cite{Nasedkin2024, JiWang2025}. 
The atmospheric compositions of exoplanets encode their accretion history and thereby provide constraints on planet formation models \cite{Oberg2011ApJ...743L..16O, Mordasini2016, Molliere2022ApJ...934...74M}.
Moderate resolution spectroscopy ($R>1,000$) can resolve the distinct spectral features of chemical species in exoplanet atmospheres \cite{Konopacky2013Sci...339.1398K}, and even detect trace elements such as isotopologues \cite{Xuan2024b}. It is a powerful way to disentangle a faint planet signal from the stellar halo especially with a space-based telescope like JWST \cite{Ruffio2024}.

We present moderate resolution spectra of the three inner planets of the HR 8799 system at mid-infrared wavelengths acquired with the JWST/NIRSpec integral field unit (IFU; R$\sim$2,700; $3$-$5\,\mu\mathrm{m}$) \cite{Boker2022}. 
The data were obtained on July 29 (UT) 2023 as part of the cycle 1 GTO program 1188 (PI: Hodapp) with a total integration time of 2.2 hours.
Observations of the HR~8799 planetary system in program 1188 were split into two sets of observations. HR~8799~b was only included in a more recent epoch acquired on July 10-11 (UT) 2024, and is not part of this analysis.
The three inner planets have planet-to-star flux ratios around $\sim2\times10^{-4}$ in the $3.8$-$5.1\,\mu\mathrm{m}$ F444W NIRCam filter \cite{Balmer2025b} making them $3-100$ times fainter than the stellar halo at their respective projected angular separation ($0.4$-$1^{\prime\prime}$). 
In order to detect the planets, we rely on the fact that the cool $\sim1000\,$K planet atmospheres have very distinct spectral features compared to the much hotter $\sim7000\,$K star. We use the \texttt{BREADS} planet detection framework developed specifically for moderate resolution spectroscopy and recently adapted for JWST/NIRSpec IFU \cite{Ruffio2024}.
The key to this approach is to perform a joint modeling of the planet and the starlight directly in the NIRSpec detector frames to limit the spectral extraction systematics that otherwise originate from the instrument's spatial undersampling \cite{Law2023}.
The resulting signal-to-noise ratio (S/N) map is shown in Figure~1 (See Methods). 
The spectral extraction of the planets relies on a custom high-pass filter to remove the starlight, which also removes the spectral continuum of the planet (See Methods). The spectrum of HR~8799~c is illustrated in Figure~2 with HR~8799~d and e shown in Extended Data Figure~1 and 2.

To measure the atmospheric compositions of HR 8799 c, d, and e, we performed Bayesian retrieval analyses using the radiative transfer tool \texttt{petitRADTRANS} \cite{molliere_petitRADTRANS_2019}. Given a set of input parameters, \texttt{petitRADTRANS} generates model spectra that we compare to the observed JWST/NIRSpec spectra with its noise covariance and $1-5~\mu$m photometry of the planets (see Methods). The $1-5~\mu$m photometry provides complementary flux information that is sensitive to cloud opacity and the thermal structure of the atmosphere. By combining the photometry with NIRSpec's moderate resolution spectra, we are able to break some degeneracies between molecular abundances, clouds, and other parameters, a limiting factor in previous analysis of emission spectroscopy of planets \cite{BrownSevilla2023}. 

The parameters of our atmospheric models include the planet mass, radius, radial velocity shift, as well as parameters describing the atmospheric composition, thermal structure, and clouds (see Methods). We primarily use chemically consistent retrievals where the composition is set by an equilibrium chemistry model from which we allow the molecular abundances to vary with the following parameters: C/O, [C/H], isotopologue ratios ($^{12}$C$^{16}$O/$^{13}$C$^{16}$O, $^{12}$C$^{16}$O/$^{12}$C$^{18}$O), H$_2$S scale factor, and quench pressure. The quench pressure accounts for carbon disequilibrium chemistry involving CO, H$_2$O, and CH$_4$ \cite{Zahnle_methane_2014}. The H$_2$S scale factor adjusts the sulfur abundance up or down from [C/H]. The enrichment of all other chemical species is assumed to scale with that of [C/H]. Fig.~2 shows that the model fits the data well within an error multiplication factor of 1.5, which we retrieved as a free parameter.

We unambiguously detect CO, H$_2$O, CH$_4$, CO$_2$, H$_2$S, $^{13}$CO and C$^{18}$O in these planets (Fig.~2). We validate the molecular detections using a second set of free retrievals, where the mass-mixing ratio of each molecule is fitted independently but assumed to be constant with pressure (see Methods). We first compare the Bayesian evidence between models that contain all molecules, and models without one specific molecule (Extended Data Table~1). Second, we confirm the molecular detections and quantify their significances using cross-correlation analysis, which yielded $5$–$28\sigma$ detections for CO$_2$, CH$_4$, and $^{13}$CO in the three planets. For H$_2$S, we obtain a $9\sigma$ detection for HR 8799 c, and a $3\sigma$ detection in HR 8799 d. C$^{18}$O is detected with $3\sigma$ confidence in planet c. 

From our retrieved molecular abundance profiles (Extended Data Fig.~3), we find the abundances of CO, CH$_4$, and H$_2$O are quenched (or set constant) from the deep atmosphere between a few to tens of bars. This indicates vigorous mixing that causes disequilibrium chemistry between these species (see Methods for details).
Our detection of CO$_2$ at moderate spectral resolution is a milestone for distant gas giants as it is an excellent tracer of metal enrichment, with an abundance that scales quadratically with atmospheric metallicity \cite{Lodders2002}. Our measurements of high CO$_2$ mass-mixing ratios ($\log{ \textrm {CO}_2}\approx-4.4$ to $-4.3$ for planets c and d) point to elevated carbon and oxygen abundances in these planets (see Extended Data Fig.~3). These results solidify the JWST/NIRCam photometric detection of CO$_2$ from ref. \cite{Balmer2025b}. From our retrievals, we measure carbon abundances of $\rm C/H=5.0^{+1.1}_{-1.4}$, $\rm C/H=9.2_{-2.4}^{+3.3}$, $\rm C/H=1.5_{-0.7}^{+1.5}$ $\times$ stellar for c, d, and e, respectively. Accounting for oxygen sequestration in silicate clouds \cite{Calamari2024}, we obtain planet C/O ratios of $0.53\pm0.02$, $0.58\pm0.01$, and $0.53^{+0.03}_{-0.04}$ (the stellar C/O is $0.59\pm0.11$, ref. \cite{Baburaj2025}), indicating a similar degree of enrichment in oxygen. These measurements reinforce the emerging trend that distant gas giants have super-stellar C/H and O/H \cite{Nasedkin2024, Balmer2024} unlike their brown-dwarf counter parts, which show stellar abundance patterns in C and O \cite{Xuan2024b, Hoch2023}. 

From the retrieved H$_2$S scale factors, we find S/H = $6.7^{+1.9}_{-1.5}$, S/H = $2.7^{+1.5}_{-1.2}$, and S/H = $2.0^{+2.4}_{-1.2} \times$ stellar for c, d, and e, respectively. Therefore, the HR 8799 c, d, and e planets have C/S ratios between $\approx1-3$ relative to their host star (Fig.~3). This indicates nearly uniform enrichment in C, O, and S across the three planets.

Although carbon and oxygen abundances have been measured in the atmospheres of many directly imaged planets \cite{Hoch2023,Xuan2024b}, these volatiles are present in both the gas and solid phase of protoplanetary disks making it difficult to unambiguously infer the accretion history of a planet from C/O and C/H alone \cite{Mordasini2016, Molliere2022ApJ...934...74M}. 
Refractory species like sulfur are better tracers of solid accretion because they remain in the solid phase throughout most of the protoplanetary disk during planet formation \cite{Kama2019, Schneider2021b, Turrini2021,Chachan2023}.
The comparably enhanced levels of sulfur, carbon, and oxygen enrichment in the atmospheres of the HR 8799 planets explicitly demonstrates that their metal enrichment is a result of efficient accretion of solids.

We can compare sulfur abundances in the wide-separation HR 8799 planets to those in short-period transiting giant planets. Sulfur was first detected in the exoplanet WASP-39b in the form of SO$_2$ \cite{Tsai_SO2}, which is a photochemical product created by intense UV irradiation from the star. Converting an SO$_2$ abundance to S/H is complicated since the production rate of SO$_2$ depends on poorly understood physical conditions and efficiencies of the relevant photochemical pathways \cite{Polman_sulfur}. On the other hand, H$_2$S is a more straightforward tracer for the overall atmospheric sulfur abundance \cite{Polman_sulfur}. H$_2$S was first detected in an exoplanet by ref. \cite{Fu2024}. They measured sub-stellar C/S, O/S, and C/O ratios for the hot Jupiter HD 189733 b, which are significantly different compared to the HR 8799 planets (Fig.~3). The H$_2$S detection in HD 189733 b was confirmed via emission spectroscopy (as opposed to transmission) with JWST/MIRI \cite{Inglis2024b}, lending credence to the measured abundances in HD 189733 b and pointing towards a true difference in atmospheric composition. The differing compositions potentially indicate that these planets experienced different formation conditions, and motivate the future study of abundance ratios in populations of exoplanets.

We use the measured abundances of C, O, and S to estimate the metal mass fraction of the planetary envelopes. Assuming that the enrichment of S reflects the enrichment of other refractory species (e.g. Fe, Si, Mg, \cite{Turrini2021, Chachan2023}), we find envelope metal mass fractions of $Z = 0.061^{+0.014}_{-0.012}$ and $0.093^{+0.029}_{-0.022}$ for HR 8799 c and d, respectively.
If the measured atmospheric metallicity is interpreted as the minimum bulk metallicity of the planet \cite{Thorngren2016}, we expect HR 8799 c and d to contain $148^{+39}_{-31}~M_\oplus$ and $272^{+86}_{-66}~M_\oplus$ of metals assuming total masses of $7.7~M_{\rm Jup}$ and $9.2~M_{\rm Jup}$, respectively \cite{Zurlo2022}. The implied bulk metallicity and heavy element mass are compatible with the mass-metallicity relationship for close-in giant planets \cite{Thorngren2016}. These metals could be accreted either via planetesimals or via pebbles that leak through imperfect traps created by growing planets \cite{Lee22,VanClepper2025}. 

The high amount of heavy element enrichment in these planets likely disfavors formation via gravitational instability, which is expected to result in these planets having elemental abundances (C/H, O/H, and S/H) similar to or slightly enhanced by up to a factor two relative to their star \cite{Helled2010, Boley2011}. 
The estimated sum of the metal mass residing within the four planets is $\sim 600~M_\oplus$ or $\sim 2~M_{\rm Jup}$, which likely comes from the dust in the disk. This is massive even for the youngest known disks, and consistent with the top 10\% percentile of the measured dust masses in Class 0/I disks, which are $<1$ Myr \cite{Tobin20}. 
In practice the required amount of metal mass will be higher, as not all solids will be accreted onto the growing planets \cite{Lin18,Chachan2021Kepler-167eSuper-Earths}. 
Given that the disk mass decreases rapidly with age \cite{Mamajek2009, Tobin20}, the large planet masses suggest that the HR 8799 planets started forming early, but the planets could have continued accreting for a longer time. 
We note that the required amount of metals could be lower if the atmospheric enrichment is larger than the deeper interior and these two regions are separated by a radiative region, as hypothesized for Jupiter to simultaneously explain its gravity data and atmospheric observations \cite{Howard2023}.
However, this compositional separation requires weak mixing in the radiative region, and its plausibility for the HR 8799 planets needs to be investigated.

Our simultaneous measurements of volatile and refractory abundances in a multi-planet system provide strong constraints on formation scenarios. The uniform enrichment of C, O, and S bears a striking resemblance to Jupiter in our solar system. In the case of Jupiter, the fact that this uniformity extends to N and noble gases suggests that the solids incorporated into Jupiter's atmosphere originated beyond the N$_2$ snowline at $\approx30$ AU \cite{Owen1999}. 
The present-day orbits of the HR 8799 planets are farther out than Jupiter, between 15 to 70 AU. However, the CO snowline is expected to be at $\sim80$ AU for the HR 8799 system given the higher luminosity of HR~8799~A compared to the Sun \cite{Chiang97}. For the HR 8799 planets, we might naturally expect to observe uniform heavy element enrichment if their envelopes were accreted beyond the CO snowline, where all the relevant species are frozen out into the solid phase and any accreted solids bring these elements in stellar proportions. However, the core formation timescale is much longer at larger orbital distances, so forming all four planets beyond 80 au is at the limit of what could reasonably be achieved under current core accretion models \cite{Johansen2017}.

The HR 8799 planets could potentially be enriched uniformly at or near their current locations within the CO snowline. However, within the CO snowline, a fraction of the C and O is expected to be in the gas phase (as CO) and the solids are not expected to contain C, O, and S in stellar proportions. To achieve close to uniform enrichment in C, O, and S within measurement uncertainties, the planets need to accrete gas and solids in a way that mostly maintains the stellar ratios in these elements. 

One possibility is that $\gtrsim 75$\% of the carbon budget is in CO$_2$ (with a snowline location of $\sim$ 10 AU; \cite{Chiang97}), or other organic molecules as opposed to CO ice. Such a carbon inventory may already exist at the protostellar stage \citep{McClure2023} or could be due to CO's conversion into CO$_2$ or other less volatile carbon compounds that remain in the solids at the planets' formation location \cite{Eistrup2018, Bosman2018}. 
Then, the sublimation of CO ice would not significantly change the composition of the solids and the accretion of such solids would lead to nearly uniform enrichment.

We note that our finding of similar enrichment in O compared to C and S for HR 8799 cde disfavors the trapping of CO into solids by water ice \cite{Lunine1985, Bar-Nun1988} as an explanation of their abundance pattern, as this would lead to a preferential O enrichment and substellar C/O ratio. 

The HR 8799 planets demonstrate that systems of multiple gas giant planets outside the solar system can exhibit a Jupiter-like enrichment pattern in both refractory and volatile species. Our results also suggest the processes leading to uniform enrichment are operating over a wide range of orbital semi-major axes in the HR 8799 system (from $15-40$ AU), and upcoming results for HR 8799~b would test whether the trend extends to 70 AU. Future observations of additional gas giants with JWST will be sensitive to sulfur as well as other refractory elements such as iron, sodium, and potassium. These observations are poised to reveal whether uniform atmospheric enrichments are a universal feature of giant planet formation at large orbital distances, thereby providing powerful constraints on formation models.

\clearpage
\backmatter

\section*{Methods}\label{sec:methods}

\subsection*{Observations}
In this work, we analyze the $3$-$5\,\mu\mathrm{m}$ JWST/NIRSpec IFU observations of HR 8799 cde in the moderate resolution spectroscopy mode ($R\sim2,700$;  filter: F290LP; grating: G395H) from cycle 1 GTO program 1188 (PI: Klaus Hodapp) obtained on July 29 (UT) 2023. 
In order to improve the spatial sampling of the instrument and better address detector-level systematics, the observations were equally split over three observatory roll angles spanning $\sim$10 degrees and each roll included twenty small dithers. 
The observatory roll angles are V3PA=$221.02^\circ$ for observation \#4, V3PA=$227.02^\circ$ for observation \#5, and V3PA=$231.01^\circ$ for observation \#6.
Each roll includes  2 detectors (NRS1 and NRS2), 2 groups, 3 integrations, and 20 dithers with NIRSpec's small cycling pattern summing up to $\sim44$ min per roll. 
The full dataset is therefore made of 120 individual detector images.

\subsection*{Raw data reduction}

We use the JWST Science Calibration Pipeline \cite{Bushouse2023} to generate calibrated detector images (\textit{*\_cal.fits}). The uncalibrated NIRSpec detector images were generated using the version 2023{\_}4a (SDP{\_}VER) of the JWST Science Data Processing (SDP) subsystem. The science calibration pipeline version ``1.12.5'' (CAL{\_}VER) stages 1 and 2 were used to produce the flux-calibrated detector images.
The version of the Calibration Reference Data System (CRDS) selection software was 11.17.19 (CRDS{\_}VER) and the CRDS context version is jwst{\_}1238.pmap (CRDS{\_}CTX) \cite{Greenfield2016}.

In general, we use the methods developed in ref.\cite{Ruffio2024} for reducing NIRSpec IFU data for high-contrast spectroscopy to detect and extract the spectrum of the planets. We focus only on the specific improvements that were made for this analysis in the following. The methods are implemented in the Python package \texttt{BREADS} \cite{breads}. 

Due to the brightness of the star, the core of the stellar PSF fully saturates up to $\sim0.15^{\prime\prime}$ at $3.0\,\mu\mathrm{m}$ ($\sim0.1^{\prime\prime}$ at $4.0\,\mu\mathrm{m}$) in the direction perpendicular to the IFU slices. 
Similarly to ref.\cite{Ruffio2024} (see Appendix C therein), we use an intermediate step between the stage 1 and 2 of the JWST pipeline to remove the 1/f correlated read noise and the transferred charges in the rate maps (\textit{*\_rate.fits}). 
The method used in ref.\cite{Ruffio2024} featured significant residuals in the IFU slices closest to the star on the detector due to the imperfect modeling of the charge transfer \cite{Boker2022} around the saturated pixels, which are not well reproduced with a spline.
This effect is more important for the HR~8799 planets due to their smaller projected separations ($0.4$-$0.9^{\prime\prime}$) compared to HD~19467~B ($1.6^{\prime\prime}$) in ref.\cite{Ruffio2024}. 

In this work, we model and subtract the spurious charges around the saturated pixels before fitting for the 1/f noise. The origin of this transfer of charges on the detector is discussed in Appendix C in  ref.\cite{Ruffio2024}. We empirically find that a Lorentzian is a good fit to the vertical profile of the additional charges on the detector.
The Lorentzian profile is built by: 1) identifying the saturated pixels in the core of the stellar PSF, 2) fitting a WebbPSF model \cite{Perrin2012SPIE.8442E..3DP,Perrin2014SPIE.9143E..3XP} to estimate the additional charges beyond the saturation threshold in the saturated pixels, 3) convolving this map of extra charges by a vertical Lorentzian profile with a full-width-at-half-maximum of 10 pixels, 4) modulating the spectral continuum of the Lorentzian model with a 5-node spline, 5) subtracting this best-fit model before applying the normal column-by-column spline model from ref.\cite{Ruffio2024} to subtract the residual 1/f noise.
We refer the reader to Section 3.3 in ref.\cite{Ruffio2024} for a definition of the spline model and nodes. The nodes define the smoothness of the spline and incidentally the number of parameters to be fitted; the sharper the spectral features, the more nodes are needed.

An important input to the subsequent modeling of NIRSpec IFU data is the sky coordinates of each pixel on the detector relative to the central star. This is referred to as the point cloud. The sky coordinates are computed from the WCS headers in the FITS file, but we apply an additional calibration to correct for the chromatic drift of the centroid of the star identified in ref.\cite{Ruffio2024} (See Fig. 9 therein).  
The centroid of the star is approximately offset by $+0.17^{\prime\prime}$ in declination compared to the WCS headers. The chromatic part of the centroid drift is of the order of $0.02^{\prime\prime}$ only. For the combined set of dithers in each roll, we fit for the centroid of a WebbPSF model on a coarse grid of wavelength: from $2.859\,\mu$m to $4.103\,\mu$m in steps of $0.005\,\mu$m for the NRS1 detector, and $4.081\,\mu$m to $5.280\,\mu$m in steps of $0.005\,\mu$m for NRS2. We then fit a second-order polynomial to the relative right ascension and declination offsets as a function of wavelength.
The best-fit polynomial is then used to recalibrate the center of the coordinate system. 

\subsubsection*{Planet detection}

Here we describe the planet detection scheme which includes the computation of signal-to-noise ratio (S/N) maps and determination of the sensitivity of the observations in terms of planet-to-star flux ratio. 

Recovering the planetary signal requires modeling and subtracting the starlight very accurately. Due to NIRSpec's spatial undersampling, ref.\cite{Ruffio2024} showed that these steps should be performed directly in calibrated detector images instead of the classical spectral cubes for NIRSpec. 
The method consists in fitting for the planet signal with a joint forward model of the planet and the starlight. 
The final data products for each detector image are three cubes for the planet flux, the planet flux uncertainties, and the planet S/N as a function of the right ascension, declination, and planet radial velocity shift.
Each one of the 120 detector images are all processed independently before being combined. 

The planet signal at each location is modeled using a \texttt{WebbPSF} model and a BTSettl atmospheric model \cite{Allard2003IAUS..211..325A} with $T_{\mathrm{eff}}=1200\,\mathrm{K}$ and $\mathrm{log}(g)=5.0$; See section 4.3 in ref.\cite{Ruffio2024} for more details. We fit for the absolute flux of the planet in the F356W filter for NRS1 and F444W for NRS2.

Following the definition in Section 4.4 in ref.\cite{Ruffio2024}, the starlight model is based on a continuum normalized spectrum of the star that is empirically derived by combining starlight in the 20 dithers of each roll after masking each planet location with a $0.16^{\prime\prime}$ radius aperture. 
We use 60 equidistant spline nodes (see Section 4.7 of ref.\cite{Ruffio2024}) for each detector for the continuum normalization of the starlight corresponding to a wavelength spacing of $\sim0.02\,\mu\mathrm{m}$. 
We also use the same prescription as ref.\cite{Ruffio2024} (Section 4.5 therein) for the principal components in the model. 

The spatial dimensions are sampled on a grid with $0.05^{\prime\prime}$ spacing from $-4^{\prime\prime}$ to $1^{\prime\prime}$ in right ascension, and $-2.5^{\prime\prime}$ to $2.5^{\prime\prime}$ in declination. 

In this work, we also fit the model over a vector of radial velocity shifts of the planet relative to the star, which are sampled every $400\,$km/s from  $-4000\,$km/s to  $4000\,$km/s. We adopt an RV of $-10.5\,$km/s for the host star \cite{Ruffio2019AJ....158..200R}.

We first combine the 60 exposures for each detector using a weighted mean of the planet fluxes (see Section 4.6 in ref.\cite{Ruffio2024}). 
In order to remove any systematics that would bias the planet fluxes in an RV-independent way, we subtract the mean flux along the RV dimension after masking the values with RV$<750\,$km/s to avoid being biased by the planet central cross-correlation peak.

We show the resulting combined flux, uncertainties, S/N map, and S/N histogram for each detector in Supplementary Figure~1. The S/N histograms are computed after masking the planets and are used to evaluate the level of systematics biasing the S/N values. 
We note that the S/N calculated with \texttt{BREADS} have a standard deviation of $1.5$ for NRS1 specifically instead of unity for NRS2. We therefore inflate the errors for NRS1 by this factor to recalibrate the S/N.

In order to combine the detection maps from the two detectors, we use the measured fluxes of HR~8799~c to convert the fluxes between the different reference filters: F356W for NRS1 to F444W for NRS2. 

The achieved detection limit is defined as 5 times the flux 1 $\sigma$ uncertainty at any position in the field of view, which is shown in Supplementary Figure~2. The sensitivity is $\sim8\times10^{-6}$ in 2.2 hours at $1^{\prime\prime}$, while is significantly higher that value of $\sim1\times10^{-6}$ predicted from ref. \cite{Ruffio2024} by only scaling the exposure time. The difference in sensitivity at this separation can be explained by the larger read noise ($\times3$) from the only reading 2 groups instead of 9, the higher planet effective temperature ($\times1.5$) and the higher level of systematics in NRS1 mentioned previously ($\times1.5$). The original motivation for choosing two groups was to minimize the effects of saturation, but these results have shown that saturation is not a significant concern and that the read noise was underestimated.

\subsubsection*{Spectral extraction}

We describe the extraction of the planet spectra and their covariance in the following.
NIRSpec's spatial undersampling makes classical subtraction of the stellar point spread function challenging, so we do not attempt to recover the spectral continuum of the planet in this work. We instead use the detector-level spline modeling described previously to fit and subtract the starlight everywhere on the detector. This step is performed without accounting for the planet signal, which means that the planet spectra will be effectively high-pass filtered, but the high-resolution spectral features will be conserved.

Here, we use fewer spline nodes (only 40 per detector row) for the spectral extraction, compared to the planet detection, to retain as much of the planet signal as possible. Indeed, more nodes means that the high-pass filtering is more aggressive. However, we place the 40 nodes in a way to minimize the starlight residuals as described below. A row on the detector is not exactly co-linear with the spectral direction. This means that the variations of the starlight continuum across a row are due to the varying wavelength as well as the varying spatial coordinates. The curvature of NIRSpec's spectral traces on the detector is stronger at the edge of the spectral band, which means that the variations of the starlight intensity are sharper on the left side of NRS1 and the right side of NRS2. Therefore, the starlight residuals will be larger at the edge of the filter for a given spacing of the nodes. These residuals were identified in ref.\cite{Ruffio2024}, but their cause was not explained. To address this issue without increasing the number of nodes, we divide the wavelength range of each detector in four equal sections and increase the node density towards the edge of the filter. For NRS1, the node spacing in each section is respectively 0.02, 0.03, 0.04, and 0.06$\,\mu$m from left to right. It is the reverse for NRS2. This spline modeling effectively creates an adaptive high-pass filtering, which is more aggressive where it is needed without unnecessarily subtracting the planet signal elsewhere.

Once the starlight has been subtracted, we interpolate the residual detector images on a regular wavelength grid following the method in ref.\cite{Ruffio2024}. We then extract the spectrum of the planets by fitting a WebbPSF model to the combined 20 dithers of each observatory roll angle at each wavelength and on a regular grid of spatial coordinates with $0.05^{\prime\prime}$ spacing. The absolute flux calibration of the WebbPSF model fit is done in the same way as ref.\cite{Ruffio2024}.
The spectral cubes from the three rolls are combined with a weighted mean and a $5\sigma$ outlier rejection to filter inconsistent flux estimates between the three rolls. The combined cube is used to extract the planet fluxes at their respective projected separation based on the relative astrometry from ref. \cite{whereistheplanet}. The spectra of each planets are shown in Fig.~2, and Extended Data Fig.~1 and 2.

One downside of this approach for spectral extraction is that the starlight is oversubtracted due to the additional flux of the planet, which is not accounted for in the model. This effect is negligible when either the planet or the speckle intensity is small compared to the other, which is the case for HR~8799~c and e.
However, a positive signal can be seen around $3.74$ and $4.0\,\mu$m for HR~8799~d in Extended Data Figure~1, which is likely a systematic artifact from the starlight subtraction as it corresponds to deep hydrogen lines of the star. The continuum of HR~8799~d represents about $10\%$ of the stellar photons at its location. Therefore, when a stellar line has a relative depth of $10\%$, we can expect residual stellar spectral features with an amplitude similar to the planet continuum. However, given the F0 spectral type of HR~8799, the stellar hydrogen lines are sparse and are not expected to affect the retrievals.

In order to estimate the noise and the covariance of each spectra, we use the spectra extracted in $0.1^{\prime\prime}$-wide annuli around the star with radii matching each planet. A disk of radius $0.3^{\prime\prime}$ is masked around each planet to avoid biasing the noise estimates. 
The flux uncertainties in each wavelength bin of the spectra is therefore computed from the sample standard deviation of the speckle spectra in each annulus.
To account for the varying spline-node spacing in each section of the wavelength range, the covariance matrix is defined as a $4\times4$ diagonal block matrix. The off-diagonal blocks are assumed to be null. Each block matrix is derived following a similar approach to \cite{Ruffio2024}.
The block correlation matrices are derived from the autocorrelation of the noise for their respective section of the spectral range and planet. The autocorrelation functions are smoothed with a 20-pixel sliding-window mean filter. They are shown in Supplementary Figure~3. The larger speckle-to-planet flux ratio for HR~8799~e leads to significantly more correlated noise in the spectra. The covariance matrix is scaled from the correlation matrix such that the diagonal of the covariance matches the vector of uncertainties of each planet spectrum.

\subsection*{Atmospheric retrieval setup}
Here, we describe the inference of atmospheric parameters from the measured planet spectra. We use the radiative transfer code \texttt{petitRADTRANS} \cite{molliere_petitRADTRANS_2019, molliere_Retrieving_2020} to generate synthetic companion templates. Our retrieval framework follows previous work on directly imaged companions \cite{Xuan2022, Xuan2024b}. For NIRSpec data, we use the line-by-line opacity sampling mode with $R=250,000$. For photometry, we use the correlated-k sampling mode with $R=100$. The physical parameters we fit for include the planet mass and radius, temperature profile, chemical abundances, and cloud structure. Following ref. \cite{Nasedkin2024}, we adopt mass priors for the planets from the stability analysis in \cite{Zurlo2022}. The fitted parameters and their priors are in summarized in Supplementary Table~1. 

\subsubsection*{NIRSpec forward model}\label{sec:fm_nirspec}
For each model spectrum from \texttt{petitRADTRANS}, we apply the following steps before comparing the model with the NIRSpec data. First, we apply a radial velocity shift to the model to align the model spectra with the data (see Radial velocities for further details). Second, we apply instrumental broadening to the model and resample the model to the data wavelengths. Specifically, we fit for the instrumental resolution as a function of wavelength ($\lambda$) using a linear relation

\begin{equation}
    R_\lambda = r_0 + r\lambda
\end{equation}

where $r$ and $r_0$ are fitted parameters. In each iteration of the fit, the model spectrum is convolved with a variable Gaussian kernel whose standard deviation $\sigma$ is given by 

\begin{equation}
    \sigma = \frac{\lambda / R_\lambda} {2\sqrt{2\log{2}}}
\end{equation}

Finally, we apply the same spline-based high-pass filtering on the model that we applied on the data. This means that we remove the best-fit spline from the model using the same node positions in the wavelength direction.
After these steps, we compute the residuals between the forward model and the data. We account for the covariance matrix in computing the log likelihood as follows

\begin{equation}
    \ln \mathcal{L} = -\frac{1}{2} \left( R^\intercal C^{-1} R + n \ln(2\pi) + \ln\det(C) \right)
\end{equation}

where $R$ is the residual array, $C$ is the covariance matrix, $n$ is the number of data points, and ${\rm det}(C)$ is the determinant of the covariance matrix. Note that $C$ has been multiplied by a factor of $e_{\rm mult}$, which is an error inflation term that we fit. 

\subsubsection*{Photometry}
In addition to the JWST/NIRSpec spectra, we also jointly fit archival photometry to anchor the continuum flux. Specifically, we use $1-2.3~\mu$m photometry from SPHERE at the Very Large Telescope \cite{Zurlo2016}, and the F356W ($\approx3.1-4.1~\mu$m) and F444W ($\approx3.8-5.1~\mu$m) photometry from JWST/NIRCam \cite{Balmer2025b}. We use the imaging F444W and F356W filter profiles to compute the synthetic photometry. The photometric error bars are inflated by $10\%$ to account for differences between the imaging and coronagraphic filters and the fact that the NIRCam data residuals are at the $10\%$ level. We add the log likelihoods from the photometry and NIRSpec components in our retrieval. 

\subsubsection*{Thermal structure}\label{sec:pt_setup}
Our baseline retrievals use the flexible pressure-temperature ($P$--$T$) parameterization from ref. \cite{Xuan2024a}, which is motivated by ref. \cite{Piette2020}. This profile is parameterized by seven $\Delta T/ \Delta P$ values between eight pressure points and the temperature at one of these pressures, $T_{\rm anchor}$. The atmospheres of each planet are modeled from $10^{1.5}$ to $10^{-7}$ bars. The selected pressure points are labeled in Supplementary Figure~4. For the radiative transfer, the eight $P$--$T$ points from our profile are interpolated onto a finer grid of 100 $P$--$T$ points using a monotonic cubic interpolation. To test the robustness of our results, we also ran a few retrievals using an alternative $P$--$T$ parametrization from ref. \cite{Zhang2023_AFLep}, which fits for temperature gradients $(d\ln{T}/d\ln{P})$ in different pressure layers. We adopted 10 pressure layers spaced logarithmically between $10^{2}$ and $10^{-7}$ bars. This alternative profile uses Gaussian priors from self-consistent atmospheric models to anchor the temperature gradients. We used the priors from ref. \cite{Zhang2025}, which derived their priors based on Sonora Diamondback models \cite{Morley2024}.

\subsubsection*{Clouds}\label{sec:cloud_choice}
We consider three different clouds models: clear, gray clouds, and the EddySed model \cite{ackerman_Precipitating_2001} as implemented in \texttt{petitRADTRANS} \cite{molliere_Retrieving_2020}. Ref. \cite{Nasedkin2024} find evidence of MgSiO$_3$ and Fe clouds in HR 8799 c, d, and e. We include these two cloud species in our EddySed models. 
While Na$_2$S and KCl cloud bases intersect our thermal profile closer to peaks in the emission contribution function (Supplementary Figure~4), we found that models with Na$_2$S or KCl clouds were not statistically favored over those with MgSiO$_3$ and Fe clouds. 

\subsubsection*{Chemistry}\label{sec:chem}
Our baseline retrieval models use the disequilibrium chemistry grid from \texttt{petitRADTRANS}, which is parametrized by C/O and [C/H] \cite{Lei2024}. The [C/H] parameter scales the abundances of all elements uniformly, while oxygen abundance is set by both [C/H] and the C/O parameter. We also fit for a H$_2$S scale factor, $f_{\rm H_2S}$, which adjusts the H$_2$S abundance up or down from the equilibrium chemistry value given each value of [C/H] and C/O (see also ref. \cite{Fu2024}). Assuming the sulfur chemistry is largely independent of the carbon chemistry, it follows that

\begin{equation}\label{eq:scale_fac}
    {\rm [S/H]} = f_{\rm H_2S} + {\rm [C/H]}
\end{equation}

Our models include line opacities from CO and its two minor isotopologues $^{13}$CO and C$^{18}$O \cite{Rothman2010}, CO$_2$ \cite{Rothman2010}, CH$_4$ \cite{Hargreaves2020}, H$_2$O \cite{Polyansky2018}, NH$_3$ \cite{Coles2019}, HCN \cite{Barber2014}, H$_2$S \cite{Azzam2016}, Na \cite{Allard2019}, K (line profiles by N. Allard, see ref.\cite{molliere_petitRADTRANS_2019}), and FeH \cite{Bernath2020}. For continuum opacities, we include the collision induced absorption (CIA) from H$_2$-H$_2$ and H$_2$-He. The reference for solar elemental abundances in our retrievals is ref. \cite{Asplund2009}. 

Additionally, we allow for carbon quenching \cite{Visscher2011} by fitting a $P_{\rm quench}$ parameter, which sets the abundances of CO, H$_2$O, and CH$_4$ at $P<P_{\rm quench}$ to be equal to their values at $P=P_{\rm quench}$ \cite{Zahnle_methane_2014}. This prescription of carbon disequilibrium does not handle CO$_2$, which ref. \cite{Beiler2024b} showed to be inaccurately captured by a quenching timescale approximation. From chemical kinetics models, ref. \cite{Beiler2024b} find that the CO$_2$ abundance is constant over pressure for most of the observable atmosphere. Therefore, in our disequilibrium chemistry models, we treat CO$_2$ separately by retrieving a constant-over-pressure abundance.

In addition to the disequilibrium chemistry models, we run several `free retrievals' where the abundance of each molecule is assumed to be constant over pressure. We include the following molecules in the baseline free retrieval: CO, $^{13}$CO, C$^{18}$O, CO$_2$, CH$_4$, H$_2$O, NH$_3$, HCN, and H$_2$S. For each planet, we perform leave-one-out experiments where we remove one molecule and re-run the free retrievals in order to validate molecule and isotopologue detections. 

\subsubsection*{Model fitting}
To sample the posteriors, we use the nested sampling package \texttt{pymultinest} \cite{Buchner2014}, a Python wrapper for \texttt{MultiNest} \cite{Feroz2019}. We adopt 1000 live points and stop sampling when the estimated contribution of the remaining prior volume to the total evidence is $<1\%$. For model comparison, we use the Bayesian evidence from each fit to calculate the Bayes factor $B$, which quantifies the relative probability of model $M_2$ compared to $M_1$ (see Supplementary Table~2).

\subsection*{Atmospheric retrieval results}

\subsubsection*{Molecular detections and non-detections}\label{sec:mole_detect}
We obtain clear detections of several molecular species in HR 8799 c, d, and e, some of which have never been detected in directly imaged planets before. In fact, even the detection of CH$_4$ in these planets has been controversial from the ground \cite{Barman2015, Ruffio2021AJ....162..290R}; here, we unambiguously identify the $\nu_3$ CH$_4$ band in all three planets. 

To validate these molecule and isotopologue detections, we run a set of free retrievals where we leave one line species out at a time. We perform these tests for CO$_2$, CH$_4$, H$_2$S, $^{13}$CO and C$^{18}$O. First, we compare the Bayesian evidence between these leave-one-out `reduced' models with the full model, which includes the full set of species. The resulting log Bayes factors are listed in Extended Data Table~1. 

While a favorable Bayes factor provides a necessary condition for molecular detection, it is not by itself a sufficient condition for detection \cite{Xuan2022}. Therefore, we perform a second test where we plot the residuals of the reduced models against a pure molecular template (Fig.~2, Extended Data Figs.~1, 2). For example, the reduced model for CO$_2$ is a model where we did not include CO$_2$ opacities, so the data residuals after subtracting this model would contain CO$_2$ lines. We quantify the CCF detection S/N by computing the cross-correlation function between the residuals CO$_2$ lines and a CO$_2$-only molecular template. These are shown as insets in Fig.~2. The same process is repeated for CH$_4$, H$_2$S, $^{13}$CO, and C$^{18}$O. 

The CCFs are computed following ref. \cite{Ruffio2017} while accounting for the inflated covariance matrices. H$_2$S and C$^{18}$O are detected with S/N of 3 in planet d, but no CCF detection is seen for planet e. An independent CCF detection of H$_2$S in planet e is not necessary to place constraints on its abundance as its existence can be safely extrapolated from the detection in planet c and d. We therefore provide the sulfur abundance measurements for all three planets in Fig.~3. 

In an analysis of low-resolution spectroscopy (near-infrared and $L$ band) and photometry of the HR 8799 planets, ref. \cite{Nasedkin2024} find their models favor the presence of HCN in HR 8799 c and e from Bayesian evidence analysis. However, they note that this detection is largely driven by low S/N, low-resolution spectroscopy from the ALES mode of LBTI/LMIRCam \cite{Doelman2022}. We do not find strong evidence of HCN in the JWST/NIRSpec spectra for any of the three planets, with $3\sigma$ log mass-mixing ratio upper limits of $\log{X_{\rm HCN}} <-5.6$ and $\log{X_{\rm HCN}}<-4.3$ for planet c and e, respectively.

\subsubsection*{CO$_2$ abundances}\label{sec:co2}

Due to model uncertainties in predicting the dis-equilibrium CO$_2$ abundance \cite{Beiler2024b}, we fitted CO$_2$ with a vertically constant mass-mixing ratio in our retrievals. We found that a model that fits the CO$_2$ abundance as vertically constant is weakly favored over a model where CO$_2$ remains in chemical equilibrium. As shown in Extended Data Fig.~3, our retrieved CO$_2$ abundance overlaps with the equilibrium CO$_2$ abundance near the peak of the emission contribution function between about $10^{0}$ to $10^{-2}$ bars, and is significantly higher than the equilibrium abundance if the C/H were stellar. Thus, our retrieved CO$_2$ abundance is consistent with the high C/H and O/H we measure for the planets. 

\subsubsection*{Isotopologue ratios}
It has been suggested that the $^{12}$CO/$^{13}$CO ratio can differ between solids and gas in the outer disk, and preferential accretion of solids or can therefore result in non-stellar values for this isotopologue \cite{Zhang2021Natur.595..370Z}. 
We measure the $^{12}$CO/$^{13}$CO  ratios in HR~8799~cde (Supplementary Table~2) and find that they are consistent at the $1-2\sigma$ level with the average value in the local interstellar medium (ISM) of $68\pm15$ \cite{ Milam2005ApJ...634.1126M}, which is also in line with recent results for directly imaged planets and brown dwarfs \cite{Zhang2024}.
In addition, we obtain $\approx3\sigma$ detections of C$^{18}$O in planets c and d. Our measurements of $^{16}$O/$^{18}$O (see Supplementary Table~2) are consistent with the average local ISM value of $557\pm30$ \cite{Wilson1999} at the $2\sigma$ level.

\subsubsection*{Evidence for clouds}
We find that the EddySed cloud models are preferred for all three planets (see Supplementary Table 2), though gray opacity models perform similarly well for planets d and e. By comparing cloudy and clear models, we find that the cloud opacity affects shorter wavelengths ($<2\mu$m) significantly more than it affects the NIRSpec wavelength range, given the locations of the MgSiO$_3$ and Fe cloud bases (see Supplementary Figure~4). Therefore, in our retrievals, the cloud parameters are mainly constrained by the VLT/SPHERE photometry.

\subsubsection*{$P$--$T$ profile, radius, and effective temperature}
The retrieved $P$--$T$ profiles for the three planets are shown in Supplementary Figure~4, along with the emission contribution functions. We checked that the $P$--$T$ profiles mostly follow the shapes of those predicted by self-consistent atmospheric models \cite{Morley2024}. For planet d, there is a slight isothermal region near 0.1 bars, which persists in retrievals using the alternative $P$--$T$ parametrization from ref. \cite{Zhang2023_AFLep}, despite the Gaussian priors imposed to prevent such behavior. This could reflect a known degeneracy between clouds and thermal structure \cite{molliere_Retrieving_2020}. Future JWST/NIRSpec data at shorter wavelengths (GO 8063; PI: Ruffio) could help better constrain the cloud parameters, and reveal whether the isothermal region persists.

From the EddySed model, we obtain $1.19\pm0.04~\Rj$, $1.09\pm0.03~\Rj$, and $0.97^{+0.10}_{-0.13}~\Rj$ for planets c, d and e. Given the stellar age of approximately 30-40 Myr \cite{Faramaz2021}, ATMO 2020 and Saumon \& Marley models \cite{Phillips2020, Saumon2008} predict that $6-9~\Mj$ planets should have radii $\approx1.2-1.35~\Rj$. Therefore, our retrieved radii for HR 8799 c, d, and e are consistent with these evolutionary models at the $1-2\sigma$ level. 

We estimate the $\lbol$ and $\Teff$ of the planets by computing $0.15-30~\mu$m low-resolution models from our chains using \texttt{petitRADTRANS}. Then, we integrate these models to calculate the bolometric luminosity and apply the Stefan-Boltzmann Law to derive $\Teff$ given the retrieved radius posteriors. We obtain $\lbollsun=-4.69\pm0.01$, $\lbollsun=-4.62\pm0.02$, and $\lbollsun=-4.64\pm0.02$ and $\Teff=1100\pm18~$K, $\Teff=1202\pm16~$K, and $\Teff=1260\pm80~$K for planets c, d, and e, respectively. Our luminosities all three planets are consistent at the $1\sigma$ level compared to estimates from ref. \cite{Nasedkin2024} using $1-5\mu$m low-resolution spectra and MIRI photometry.

\subsubsection*{Disequilibrium chemistry and vertical mixing}
From our baseline EddySed models, we constrain the quench pressure to $24_{-7}^{+5}$, $17_{-8}^{+9}$, and $14_{-12}^{+10}$ bars for planets c, d, and e respectively. These values correspond to quenching below the observed photosphere, indicating that the molecular abundances of CO, CH$_4$ and H$_2$O are homogenized or vertically constant in the pressure range over which our data are sensitive. By definition, the quench pressure is the point where the chemical timescale ($\tau_{\rm chem}$) of a reaction equals the mixing timescale \cite{Zahnle_methane_2014}. The mixing timescale ($\tau_{\rm mix}$) is 

\begin{equation}
    \tau_{\rm mix} = L^2 / K_{\rm zz} 
\end{equation}

where $L$ is the length scale. In general, $L=\alpha H$, where $H$ is the pressure scale height and $\alpha$ is a scaling factor. ref. \cite{smith_Estimation_1998} found that $L\approx0.1H$ for several reactions in the atmospheres of Jupiter and Neptune. We set $\tau_{\rm mix} = \tau_{\rm chem}$ at $P=P_{\rm quench}$ to estimate $K_{\rm zz}$. To do so, we first compute $\tau_{\rm chem}$ from eq. 12-14 in ref. \cite{Zahnle_methane_2014} using our posterior chain for $P_{\rm quench}$, the temperature at $P_{\rm quench}$ ($T_{\rm quench}$), and the atmospheric metallicity. Second, we compute the pressure scale height $H = \frac{k_B T}{\mu\, m\, g}$ 
($k_B$: Boltzmann constant; $\mu$: mean molecular weight in amu; $m$: atomic mass unit; $g$: surface gravity), where $\mu$ is computed from our retrieved molecular abundance profiles, and $g$ is from our retrieved mass and radius. This approach yields $\log (K_{\rm zz} / \rm{cm^2 s^{-1}})= 11.8^{+0.5}_{-0.7}$, $12.1^{+0.9}_{-1.3}$, and $10.0^{+2.3}_{-4.3}$ for planets c, d, and e, respectively.  

Our estimates of $K_{\rm zz}$ above are at or above the values predicted by mixing length theory \cite{Gierasch_convect1985}, which in the case of full convection, yield upper limits of $\log (K_{\rm zz} / \rm{cm^2 s^{-1}}) \approx10$. In reality, $K_{\rm zz}$ is not constant with altitude \cite{Mukherjee2022}, and using the temperature at $P_{\rm quench}$ to compute $K_{\rm zz}$ may result in unphysically high values \cite{Nasedkin2024}. To address this, ref. \cite{Nasedkin2024} chose to use the planet $\Teff$ as a representative temperature to replace $T_{\rm quench}$ in the above calculation. Since $K_{\rm zz}$ is exponential in T, the choice of T has a substantial impact on the results. If using $\Teff$, we would obtain much smaller $\log (K_{\rm zz}) \approx2-4$ for all three planets. This highlights the uncertainty in estimating $K_{\rm zz}$ from atmospheric retrievals. Additional modeling work is required to provide a framework that allows non-constant $K_{\rm zz}$ in the atmosphere, and thereby infer more accurate $K_{\rm zz}$ values.

\subsubsection*{Radial velocities}
 
The systematic wavelength calibration uncertainty for JWST/NIRSpec G395H spectrum can be as large as $\approx20$km/s (\url{https://jwst-docs.stsci.edu/jwst-calibration-status/nirspec-calibration-status/nirspec-ifu-calibration-status}). Our goal of fitting the planetary RV is to re-calibrate the imperfect wavelength solution such that the models align with the data, rather than to obtain scientifically valuable constraints. We find barycentric-corrected RVs of $-16.5\pm0.6~$km/s, $-19.3\pm0.8~$km/s, and $-17.3\pm1.6~$km/s for HR 8799 c, d, e respectively. Given the large systematic uncertainty, these values are consistent with orbital predictions of planetary RV$~\approx-10-13$km/s \cite{Ruffio2021AJ....162..290R}.

\subsection*{The chemical abundances of HR 8799 A}\label{sec:stellar_abund}
HR 8799 A is a $\lambda$ Boo type chemically peculiar star, which is characterized by a deficiency of iron-peak elements such as Fe, Ni, and Mn. The origin of $\lambda$ Boo stars is a matter of active debate, but possibilities include the accretion of gas from a circumstellar object or diffuse ISM cloud \cite{Jura2015} or embedded planets that deplete the dust \cite{Kama2015}. Lighter elements such as C, O, N, and S are observed to have near-solar abundances in $\lambda$ Boo stars \cite{Kamp2001}. Therefore, in this paper we assume the measured photospheric S, C, and O abundances of HR 8799 A \cite{Baburaj2025} represent the original abundances of the planet-forming environment. 

\subsubsection*{Data availability} 
The NIRSpec data used in this paper are from JWST GTO program 1188 (PI Hodapp) and are publicly available \cite{dataset:NIRSpecHR8799} from the Mikulski Archive for Space Telescopes (MAST). The reduced NIRSpec planet spectra are available on Zenodo\cite{ruffioXuan_2025_17536808_Zenodo}.
The NIRCam photometry are taken from ref. \cite{Balmer2025b}, and the archival VLT/SPHERE photometry are obtained from ref. \cite{Zurlo2016}.

\subsubsection*{Code availability} 
The NIRSpec data reduction was performed using the BREADS package at \url{https://breads.readthedocs.io/} \cite{breads}. The atmospheric models were generated using the \texttt{petitRADTRANS} radiative transfer tool available at \url{https://petitradtrans.readthedocs.io/}.

\backmatter

\noindent \textbf{Correspondence and requests for materials} should be addressed to Jean-Baptiste Ruffio.

\bmhead{Acknowledgments}
We thank Evert Nasedkin, Yapeng Zhang, Julie Inglis, Nicole Wallack, Paul Mollière, Tamara Molyarova, Kazumasa Ohno, Jonathan Fortney, Ruth Murray-Clay, Mark Wyatt, Geoffrey Blake, Samuel Beiler, Björn Benneke, Hilke Schlichting, Jens Kammerer, Stephan Birkmann, and Nora Lützgendorf for helpful discussions. We thank Guangwei Fu for sharing abundance values for HD 189733 b, and Luis Welbanks for discussions on Sulfur measurements for transiting planets.  Part of this work was supported by the National Aeronautics and Space Administration under Grants/Contracts/Agreements No. 80NSSC25K7300 (J.-B.R.) issued through the Astrophysics Division of the Science Mission Directorate. Any opinions, findings, and conclusions or recommendations expressed in this work are those of the author(s) and do not necessarily reflect the views of the National Aeronautics and Space Administration.
J.W.X. acknowledges support for this work through the NASA FINESST Fellowship award 80NSSC23K1434. Part of this work was carried out at the Jet Propulsion Laboratory, California Institute of Technology, under a contract with the National Aeronautics and Space Administration (80NM0018D0004). This work is based on observations made with the NASA/ESA/CSA James Webb Space Telescope. The data were obtained from the Mikulski Archive for Space Telescopes at the Space Telescope Science Institute, which is operated by the Association of Universities for Research in Astronomy, Inc., under NASA contract NAS 5-03127 for JWST. These observations are associated with program 1188.

\bmhead{Author Contribution}
J-B.R. led the data reduction and coordinated the analyses, J.W.X. led the modeling analysis with atmospheric retrievals. J-B.R and J.W.X. co-wrote the manuscript with contributions from Y.C., E.J.L., and A.K. Y.C. and E.J.L. led the planet formation interpretation. W.O.B. and G.B. analyzed and provided the NIRCam photometry data. K.H. designed the JWST/NIRSpec GTO observing program. C.B, Q.K., H.K., D.M., and M.P. provided advice throughout the project. M.Y. helped with the design and preparation of the observations. All authors commented on the manuscript.

\noindent \textbf{Competing interests} The authors declare no conflict of interests.

\begin{figure}[h]
\centering
\includegraphics[trim={0cm 1.5cm 0cm 0.5cm},clip,width=0.5\linewidth]{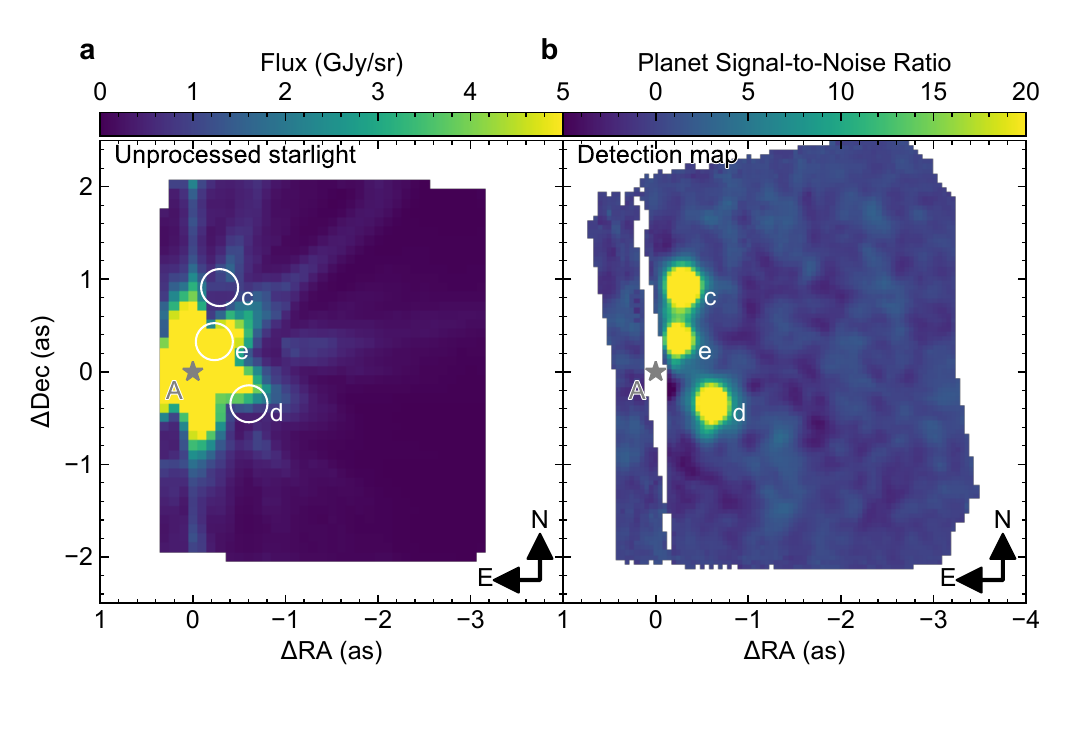}
\caption{ Detection of the three inner planets c, d, and e orbiting the star HR~8799 with the moderate resolution mode of JWST/NIRSpec IFU in the $3-5\,\mu$m spectral range.
\textbf{a.} Median spectral cube prior to starlight subtraction using the standard JWST pipeline reduction.
\textbf{b.} Signal-to-noise ratio map for planet detection. HR~8799~c, d, and e are detected with an S/N of 119, 94, 67, respectively. }\label{fig:summarydetec}
\end{figure}

\begin{figure}[h]
\centering
\includegraphics[trim={0cm 0cm 0cm 0cm},clip,width=1\linewidth]{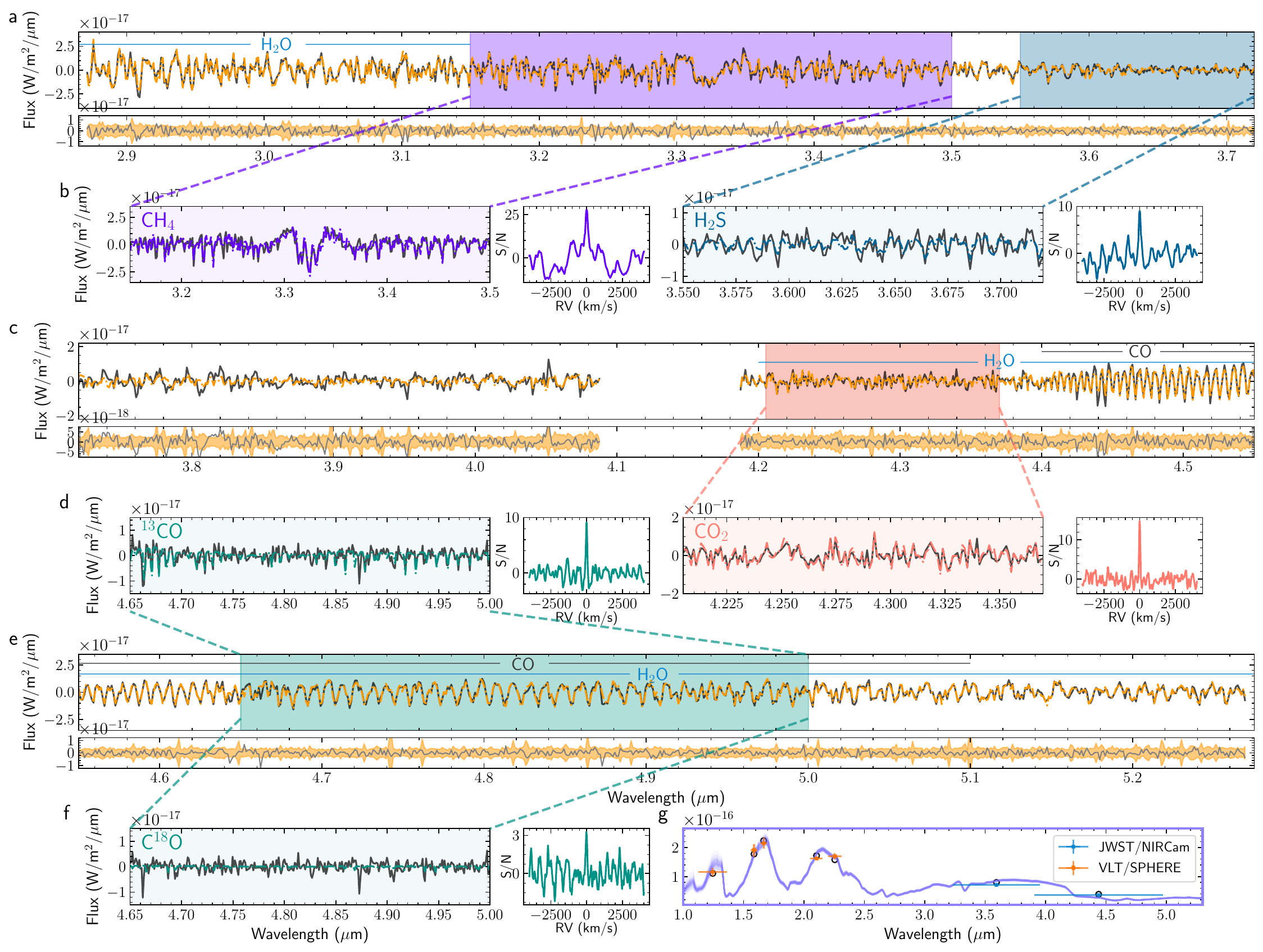}
\caption{JWST/NIRSpec spectrum of HR 8799 c. Panels a, c, e show the observed spectrum in black and the best-fit model in orange. In the sub-panels below, the residuals are plotted as gray lines and the $1.5\sigma$ uncertainties are shown as orange contours. The factor of 1.5 comes from the retrieved error scaling factor. Panels b, d, and f show data residuals after fitting an atmospheric model without a given species (CH$_4$, H$_2$S, CO$_2$, $^{13}$CO, C$^{18}$O) in black, and the corresponding molecular templates in color. The similarity between the data residuals and molecular templates indicate that the highlighted species contribute significantly to the planet's spectra. On the right insets, we plot the cross-correlation functions (CCF) between the data residuals and models in the left insets. The CCF provides an estimate of the detection S/N for each molecule. Panel g shows the photometry data in blue and orange points (with error bars representing $1\sigma$ uncertainties in y, and photometric bandpass sizes in x), the best-fit photometry model in open circles, and random draws of the model spectrum at $R=100$ in purple.
}\label{fig:spec}
\end{figure}

\begin{figure}[h]
\centering
\includegraphics[trim={0cm 0cm 0cm 0cm},clip,width=0.7\linewidth]{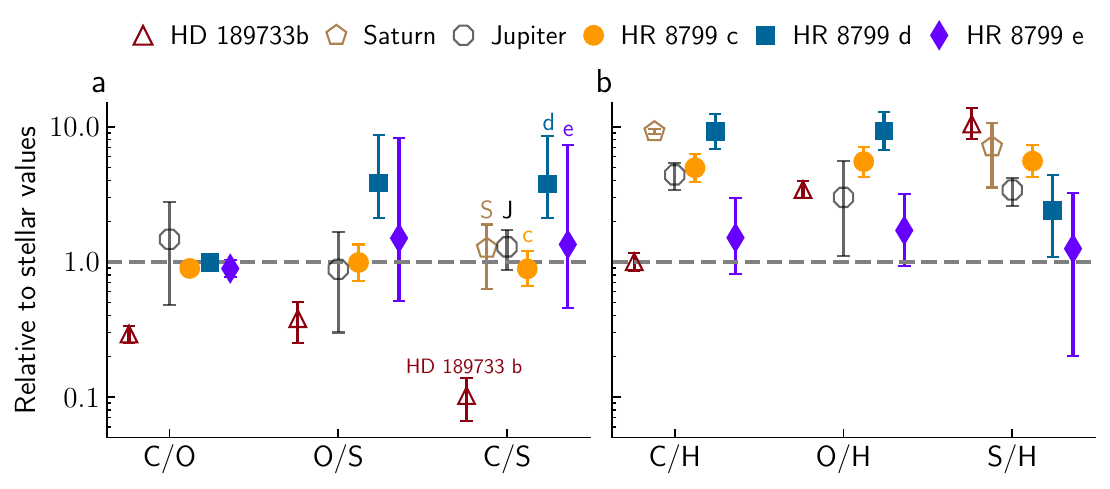}
\caption{Measured atmospheric elemental abundances of HR 8799 c, d, and e. \textbf{a}, the elemental abundance ratios between carbon, oxygen, and sulfur. \textbf{b}, the C, O, and S abundances compared to hydrogen. For all points, the y error bars represents the $1\sigma$ uncertainty on the abundance value. For comparison, we overplot measured values for Jupiter and Saturn compiled in ref. \cite{Guillot2023}, and the transiting hot Jupiter HD 189733 b \cite{Fu2024}. Each ratio is normalized to the respective host star values, such that values above and below 1 represent relative enhancement and depletion compared to the star. The HR 8799 planets show enriched atmospheric C, O, and S abundances, while the relative amounts of heavy element enrichment (C/O, O/S, C/S) are consistent with the stellar value. Such a trend is also observed for Jupiter and Saturn, although Saturn's S/H measurement remains tentative \cite{Atreya2018}. The stellar abundances of HR 8799 A are taken from ref. \cite{Baburaj2025}.}\label{fig:abundances}
\end{figure}

\hfill \pagebreak
\setcounter{page}{1}
\setcounter{figure}{0}
\setcounter{table}{0}
\renewcommand{\figurename}{Extended Data Fig.}
\renewcommand{\tablename}{Extended Data Table}

\begin{figure}[h]
\centering
\includegraphics[trim={0cm 0cm 0cm 0cm},clip,width=1\linewidth]{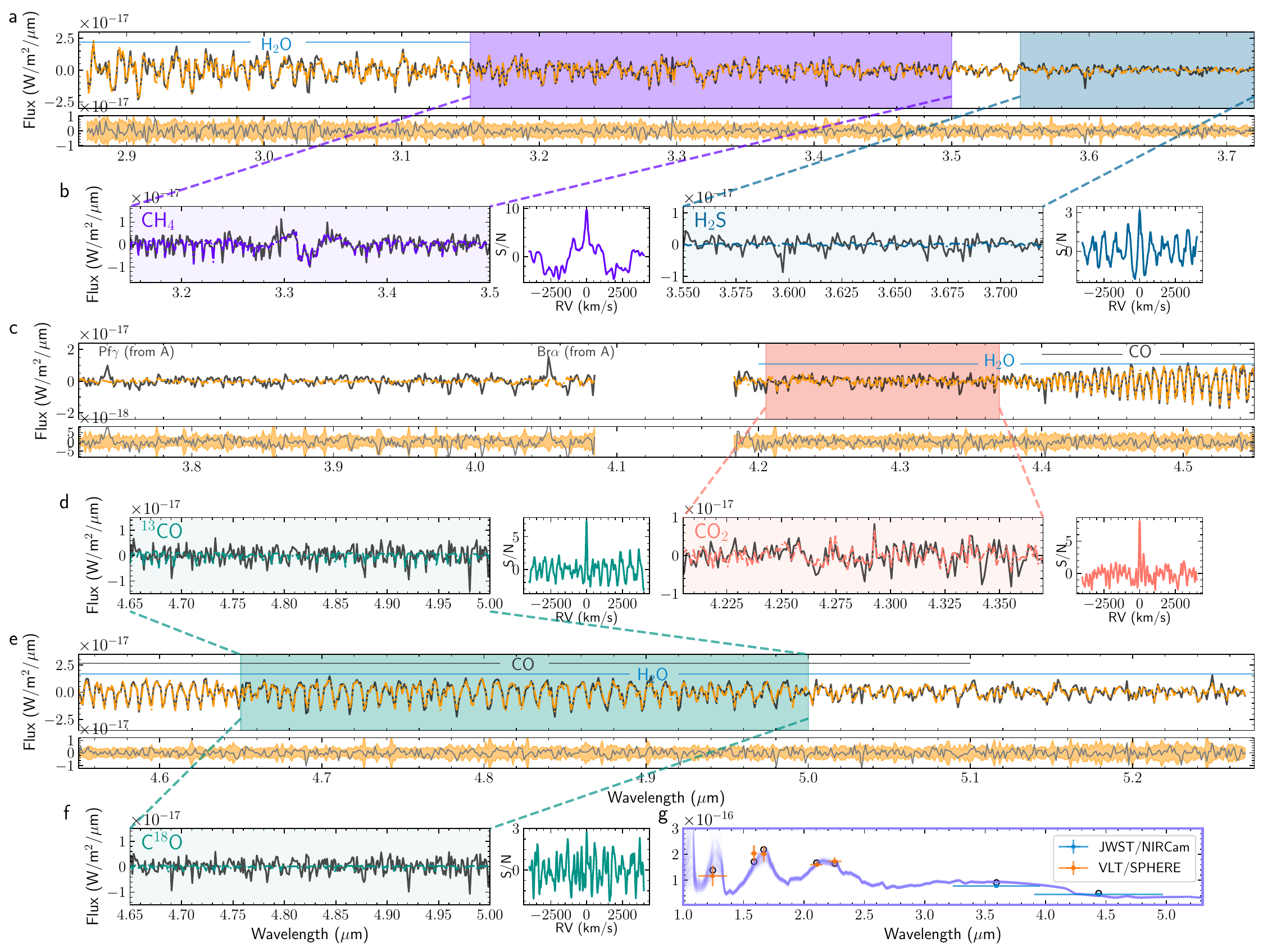}
\caption{JWST/NIRSpec spectrum of HR 8799 d. See caption of Figure~\ref{fig:spec}. Panels a, c, e show the observed spectrum in black and the best-fit model in orange. In the sub-panels below, the residuals are plotted as gray lines and the $1.5\sigma$ uncertainties are shown as orange contours. 
The factor of 1.5 comes from the retrieved error scaling factor. We indicate locations of positive residuals in panel c, which result from over-subtraction of Br$\alpha$ and Pf$\gamma$ lines in HR 8799 A (see Spectral Extraction). 
Panels b, d, and f show data residuals after fitting an atmospheric model without a given species (CH$_4$, H$_2$S, CO$_2$, $^{13}$CO, C$^{18}$O) in black, and the corresponding molecular templates in color. 
The similarity between the data residuals and molecular templates indicate that the highlighted species contribute significantly to the planet's spectra. On the right insets, we plot the cross-correlation functions (CCF) between the data residuals and models in the left insets. The CCF provides an estimate of the detection S/N for each molecule. Panel g shows the photometry data in blue and orange points (with error bars representing $1\sigma$ uncertainties in y, and photometric bandpass sizes in x), the best-fit photometry model in open circles, and random draws of the model spectrum at $R=100$ in purple.}\label{fig:spec_d}
\end{figure}
\hfill \pagebreak

\begin{figure}[h]
\centering
\includegraphics[trim={0cm 0cm 0cm 0cm},clip,width=1\linewidth]{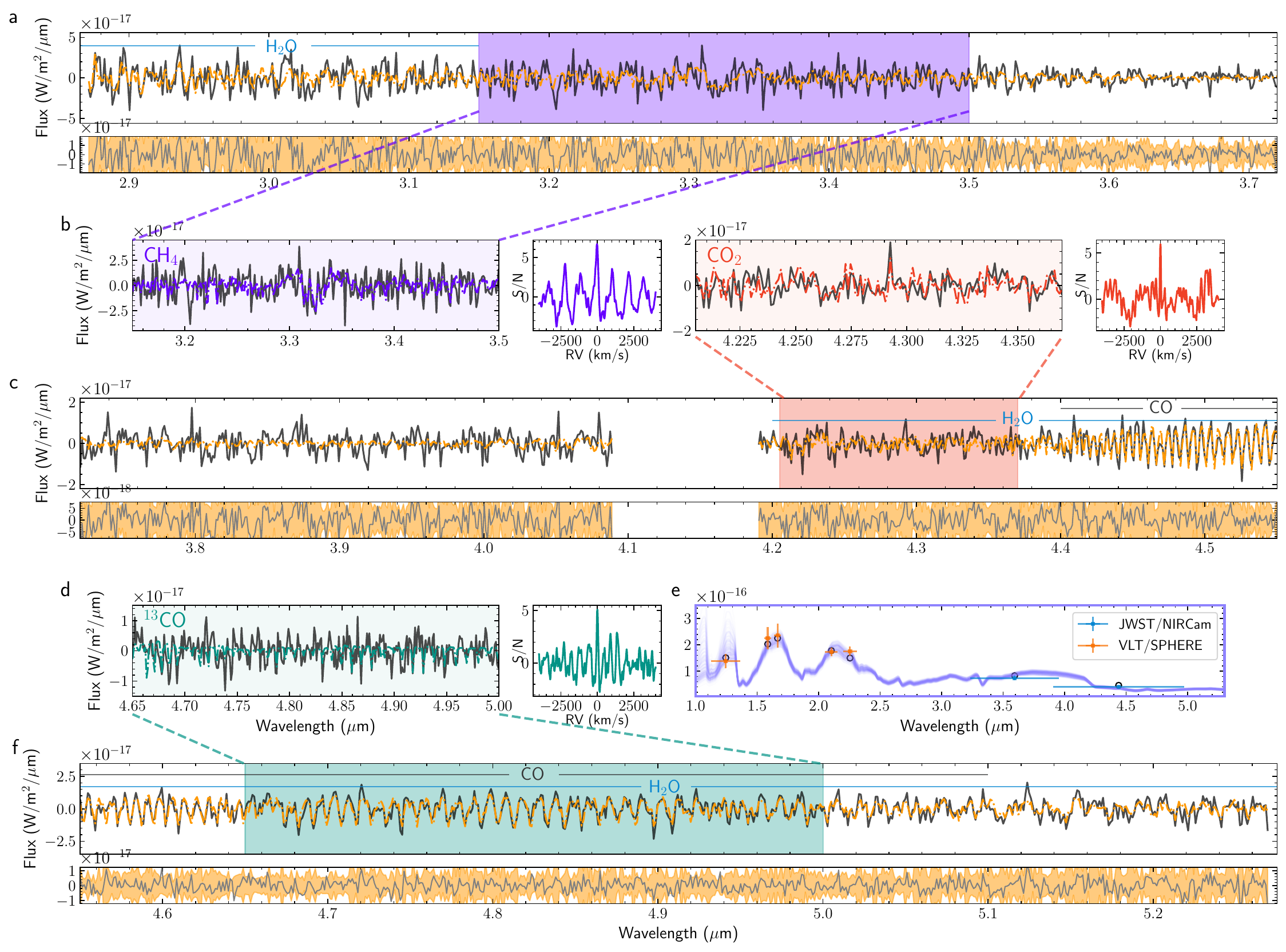}
\caption{JWST/NIRSpec spectrum of HR 8799 e. Panels a, c, f show the observed spectrum in black and the best-fit model in orange. In the sub-panels below, the residuals are plotted as gray lines and the $1.5\sigma$ uncertainties are shown as orange contours. The factor of 1.5 comes from the retrieved error scaling factor. Panels b and d show data residuals after fitting an atmospheric model without a given species (CH$_4$, CO$_2$, $^{13}$CO) in black, and the corresponding molecular templates in color. The similarity between the data residuals and molecular templates indicate that the highlighted species contribute significantly to the planet's spectra. On the right insets, we plot the cross-correlation functions (CCF) between the data residuals and models in the left insets. The CCF provides an estimate of the detection S/N for each molecule. Panel e shows the photometry data in blue and orange points (with error bars representing $1\sigma$ uncertainties in y, and photometric bandpass sizes in x), the best-fit photometry model in open circles, and random draws of the model spectrum at $R=100$ in purple.}\label{fig:spec_e}
\end{figure}
\hfill \pagebreak

\begin{table}[t!]
    \centering
    \caption{Comparison of Free Retrievals with Vertically-Constant Abundances}
    \label{tab:bayes_free}
    \small
    \begin{tabular}{ccccc}
        \hline
        Planet & Molecule & $\Delta$ln($B$) & CCF S/N \\ 
        \hline
        c & CO$_2$ & -270.4 & 14.8 \\
        & H$_2$S & -87.8 & 9.0 \\ 
        & CH$_4$ & -727.1 & 27.7 \\
        & $^{13}$CO & -103.3 & 9.1 \\
        & C$^{18}$O & -4.8 & 3.3 \\
        \hline
        d & CO$_2$ & -85.1 & 8.1 \\
        & H$_2$S & -3.1 & 3.2 \\
        & CH$_4$ & -100.4 & 9.7 \\
        & $^{13}$CO & -54.4 & 7.4 \\
        & C$^{18}$O & -4.9 & 2.8 \\
        \hline
        e & CO$_2$ & -52.7 & 5.9 \\
        & H$_2$S & -3.8 & -- \\
        & CH$_4$ & -48.3 & 6.7\\
        & $^{13}$CO & -28.6 & 5.1 \\
        & C$^{18}$O & -0.8 & -- \\
        \hline
    \end{tabular}
    \begin{flushleft}
        For each planet, \textbf{the log Bayes factor ($\Delta$ln$B$) difference} is calculated by comparing a reduced model without a given molecule (CO$_2$, H$_2$S, CH$_4$, $^{13}$CO, C$^{18}$O) to the full model. Negative Bayes factors indicate that the reduced model is disfavored. In addition, we list the cross-correlation function (CCF) S/N for each molecule. A horizontal strike indicates the molecule is not detected to $>2\sigma$ significance in the planet. 
    \end{flushleft}
\end{table}

\begin{figure}
    \centering
    \includegraphics[width=0.7\linewidth]{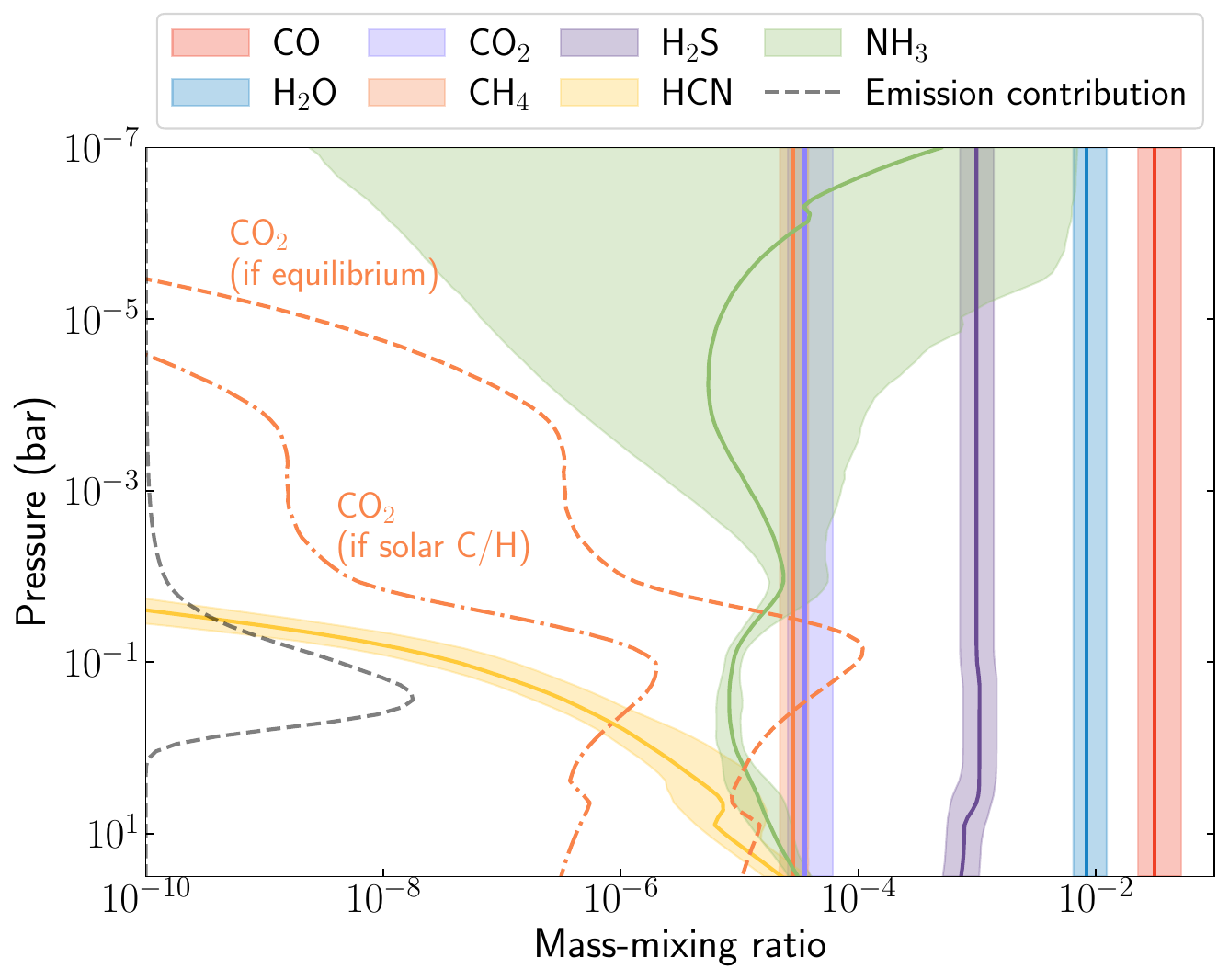}
    \caption{Mass-mixing ratios for major molecular species included in our baseline retrieval (dis-equilibrium chemistry, EddySed cloud model) of HR 8799 c. The shaded regions represent $2\sigma$ (95\%) confidence intervals, while the solid lines represent the median values. For reasons noted in Methods, we fit CO$_2$ with a vertically constant abundance profile (orange). We overplot the predicted chemical equilibrium CO$_2$ abundance from our best-fit model as the dashed orange line, whereas the dashdot orange line shows the equivalent abundance if the atmosphere instead has a solar C/H value. The wavelength-weighted emission contribution function for the NIRSpec model ($2.85-5.3~\mu$m) is shown as the dashed gray line.}
    \label{fig:mmrs_co2}
\end{figure}
\hfill \pagebreak

\hfill \pagebreak
\setcounter{page}{1}
\setcounter{figure}{0}
\setcounter{table}{0}
\renewcommand{\figurename}{Supplementary Fig.}
\renewcommand{\tablename}{Supplementary Table}

\begin{figure*}
  \centering
\includegraphics[trim={0cm 0.0cm 0cm 0.0cm},clip,width=1\linewidth]{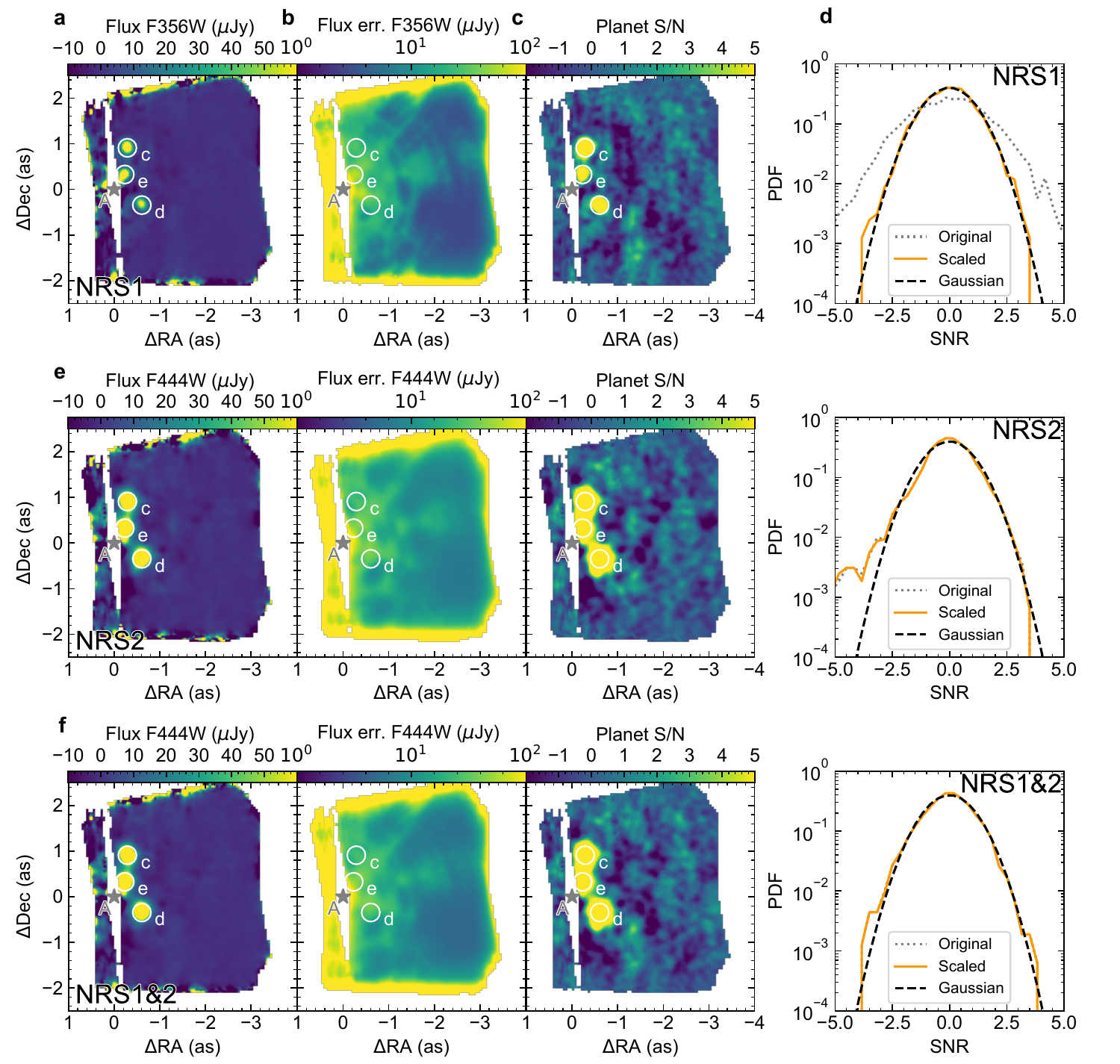}
  \caption{ Planet detection using the \texttt{BREADS} forward model framework for JWST/NIRSpec IFU.
  \textbf{a.} Planet absolute flux from fitting NRS1. The flux is expressed in the F356W filter but the fits include the entire NRS1 spectral range.
  \textbf{b.} Corresponding flux uncertainty for NRS1. 
  \textbf{c.} Planet signal-to-noise ratio (S/N) map for NRS1.
  \textbf{d.} Histogram of the (S/N) map after masking the planets. Comparing the S/N histogram to a Gaussian distribution ensures the validity of the planet detection limits.
  \textbf{e.} is the same for NRS2, while \textbf{f.} results from combining NRS1 and NRS2 together. These panels are made similarly to Figure 13 and Section 4.6 in \cite{Ruffio2024}.
  }
  \label{fig:detecmaps}
\end{figure*}

\begin{figure}[h]
\centering
\includegraphics[trim={0cm 0.0cm 0cm 0.0cm},clip,width=0.5\linewidth]{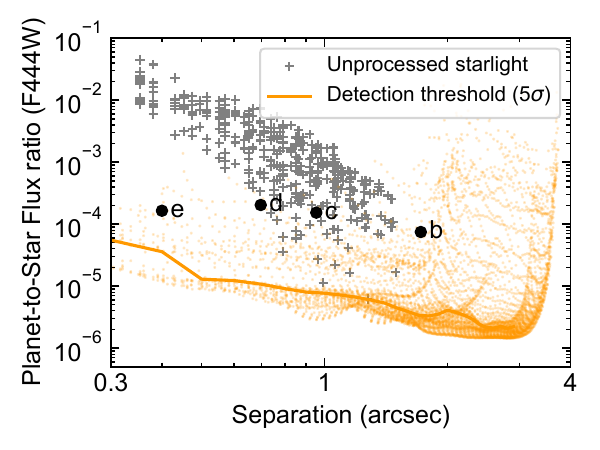}\caption{Planet $5\sigma$ detection limits for the fully combined dataset defined in the F444W filter. We use the estimated planet-to-star flux ratios and reference stellar fluxes from ref. \cite{Balmer2025b}. Each point represents a single spaxel in the field of view. The median sensitivity is calculated in concentric annuli that are $0.1^{\prime\prime}$ wide.
}\label{fig:contrast}
\end{figure}
\hfill

\begin{figure*}
  \centering
  \includegraphics[trim={0cm 0cm 0cm 0cm},clip,width=1\linewidth]{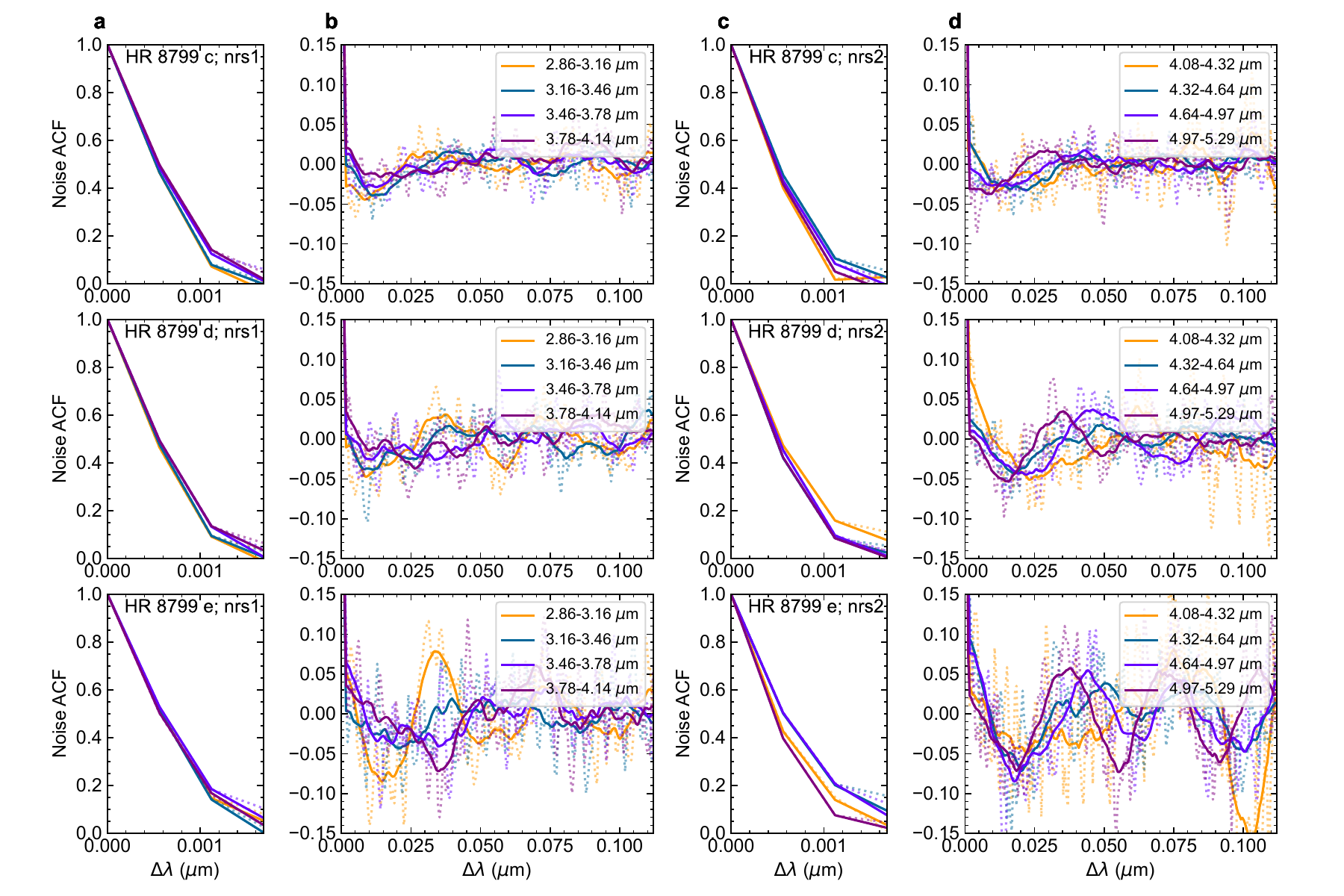}
  \caption{Autocorrelation functions of the noise for each planet spectra.
  \textbf{a.} Small-scale autocorrelation of each planet in NRS1. The small-scale correlation stems from the interpolation of the detector images on a regular wavelength grid. 
  \textbf{b.} features the large-scale autocorrelation of each planet in NRS1. The large-scale correlation stems from the residual starlight caused by the curvature of the IFU spectral traces on the NIRSpec detectors and the limited number of spline nodes. The spectral scale of the auto-correlation matches the spline node spacing. 
  \textbf{c.} and \textbf{d.} are the NRS2 equivalent of \textbf{a.} and \textbf{b.}
  }
  \label{fig:autocorr}
\end{figure*}

\begin{table*}
    \centering
    \caption{Fitted Parameters and Priors for Retrievals}
    \label{tab:param_prior}
    \small
    \begin{tabular}{ll|ll}
        \hline
        Parameter & Prior & Parameter & Prior \\
        \hline
        Mass ($\Mj$) & $\mathcal{N}(\mu_{\rm M, dyn}, \sigma_{\rm M, dyn})^{\rm (a)}$ & Radius ($\Rj$) & $\mathcal{U}(0.6, 2.0)$  \\
        $T_{\rm anchor}$ [log($P$)=-0.5]$^{\rm (b)}$ (K) & $\mathcal{U}(500, 1500)$ & RV (\kms) & $\mathcal{U}(-50 , 50)$ \\
        $\Delta T_1$ [1.5 to 0.5]$^{\rm (b)}$ (K) & $\mathcal{U}(500, 1500)$ & $\rm C/O$ & $\mathcal{U}(0.1,1.6)$ \\
        $\Delta T_2$ [0.5 to 0.0]$^{\rm (b)}$ (K) & $\mathcal{U}(0, 500)$ & $\rm [C/H]$ & $\mathcal{U}(-1.0,2.0)$ \\
        $\Delta T_3$ [0.0 to -0.5]$^{\rm (b)}$ (K) & $\mathcal{U}(0, 500)$ & \logco & $\mathcal{U}(0, 8)$  \\
        $\Delta T_4$ [-0.5 to -1.0]$^{\rm (b)}$ (K) & $\mathcal{U}(0, 400)$ & \logcoo & $\mathcal{U}(0, 8)$ \\
        $\Delta T_5$ [-1.0 to -2.0]$^{\rm (b)}$ (K) & $\mathcal{U}(0, 400)$ & $\log({f_{\rm{sed}}})^{\rm (e)}$ & $\mathcal{U}(-5, 1)$ \\
        $\Delta T_6$ [-2.0 to -4.0]$^{\rm (b)}$ (K) & $\mathcal{U}(0, 400)$ & ${\rm log}(K_{\rm zz}/\rm{cm^2 s^{-1}})^{\rm (e)}$ & $\mathcal{U}(2, 15)$ \\ 
        $\Delta T_7$ [-4.0 to -7.0]$^{\rm (b)}$ (K) & $\mathcal{U}(0, 400)$  & $\sigma_{\rm g}^{\rm (e)}$ & $\mathcal{U}(1.05, 3)$ \\ 
        log(gray opacity/$\rm{cm}^2 g^{-1}$)$^{\rm (c)}$ & $\mathcal{U}(-6, 0)$  &  ${\rm log}(X_{\rm cloud})^{\rm (e)}$ & $\mathcal{U}(-8.0, 0.0)$ \\
        Error multiple$^{\rm (d)}$ & $\mathcal{U}(1, 5)$ & log(H$_2$S scale factor) & $\mathcal{U}(-3, 2)$ \\
        log(CO$_2$) mass-mixing ratio & $\mathcal{U}(-8, -2)$ & log($P_{\rm quench}$ / bars) & $\mathcal{U}(-5.0, 2.0)$ \\
        $r_0$ & $\mathcal{U}(-650, 1400)$ & $r$ & $\mathcal{U}(200, 1300)$ \\
        \hline
    \end{tabular}
    \begin{flushleft}
    $\mathcal{U}$ stands for a uniform distribution, with two numbers representing the lower and upper boundaries. $\mathcal{N}$ stands for a Gaussian distribution, with numbers representing the mean and standard deviation. The $P$--$T$ parameters are described and the cloud parameters are described in Methods. All logarithmic values in this table are base 10. \\
    $^{\rm (a)}$ We use mass priors from the stability analysis in ref. \cite{Zurlo2022}. The adopted values are $7.7\pm0.7~\Mj$, $9.2\pm0.7~\Mj$, and $7.6\pm0.9~\Mj$ for planets c, d, and e respectively. \\
    $^{\rm (b)}$ The pressure at $T_{\rm anchor}$ and pressure points between which we fit $\Delta T$ values in our $P$--$T$ profile are given in square brackets. They are in log(bar) units.\\
    $^{\rm (c)}$ Parameter for the gray opacity cloud model.\\
    $^{\rm (d)}$ We multiply this term to the NIRSpec data covariance matrix to account for underestimation of the uncertainties.\\
    $^{\rm (e)}$ Parameters for the EddySed cloud model. ${\rm log}(X_{\rm cloud})$ refers to the log cloud mass fraction at the cloud base. The same prior is used for MgSiO$_3$ and Fe, the two cloud species in our EddySed model.
    \end{flushleft}
\end{table*}
\hfill \pagebreak

\begin{figure}
    \centering
    \includegraphics[width=0.6\linewidth]{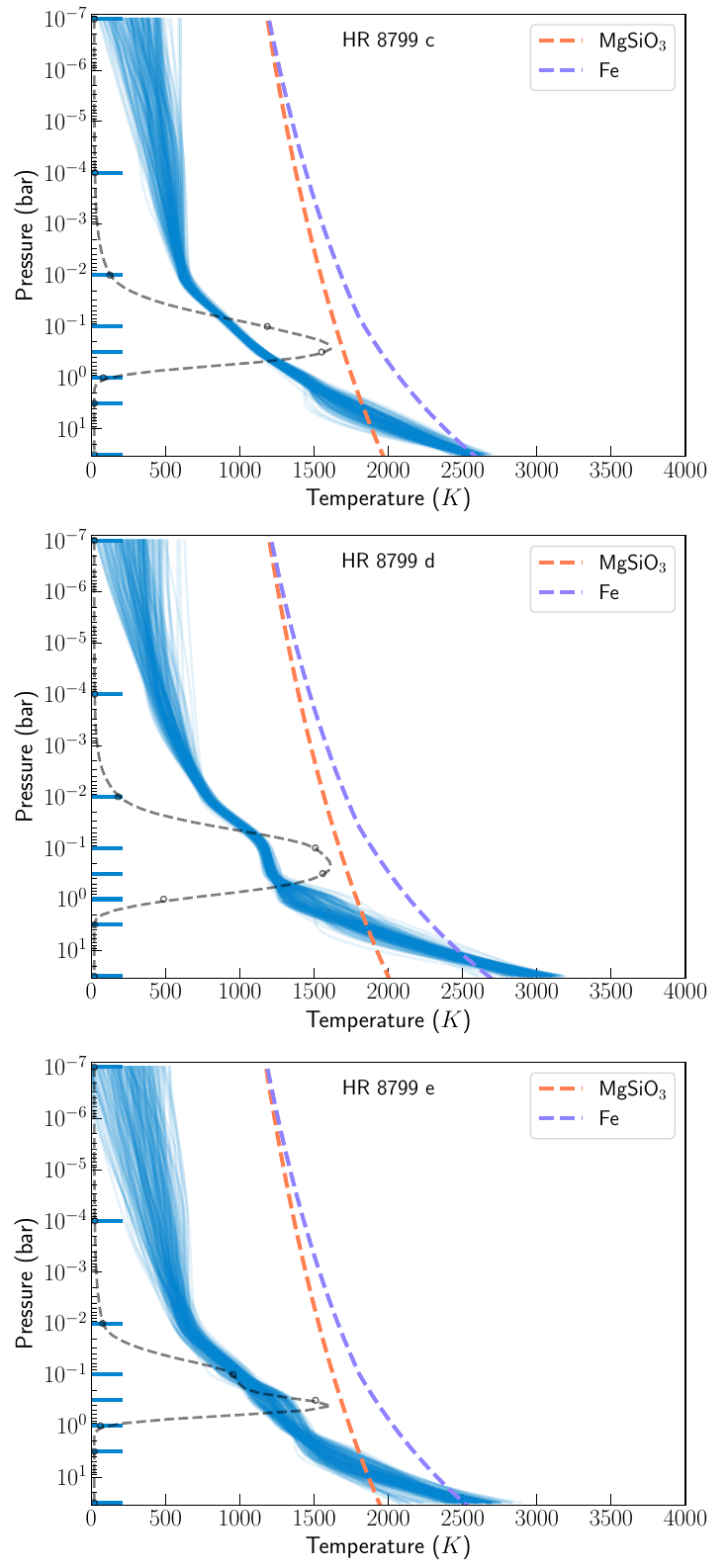}
    \caption{Top panel: random draws of the retrieved $P$--$T$ profiles for HR 8799 c in blue. The MgSiO$_3$ and Fe cloud condensation curves are plotted as colored dashed lines. The horizontal blue lines mark pressure points between which we fit $\Delta T$ values in the $P$--$T$ parameterization. The dashed gray curve shows the wavelength-weighted emission contribution function for the JWST/NIRSpec model  ($2.85-5.3~\mu$m). Middle and bottom panels are equivalent plots for HR 8799 d and e.}
    \label{fig:pt_c}
\end{figure}
\hfill \pagebreak

\clearpage

\begin{table*}[t!]
    \centering
    \caption{\textbf{Results of Spectral Retrievals for HR 8799 c, d, e.}} 
    \label{table:spec_results}
    \resizebox{\textwidth}{!}{ 
    \begin{tabular}{cc|cccccccc|c}
        \toprule
        Object & Model & Radius & C/O & C/H ($\times$ stellar) & S/H ($\times$ stellar) & \co & \coo & $T_\textrm{eff}$ & $P_{\rm quench}$ & ln($B$) \\
         &  & ($\Rj$) & & & & & & (K) & & \\
        \midrule
        HR 8799c & EddySed & $1.19\pm0.04$ & $0.53\pm0.02$ & $5.0^{+1.1}_{-1.4}$  & $5.6_{-1.3}^{+1.7}$ & $66_{-11}^{+13}$ & $2500_{-1300}^{+5000}$ & $1110\pm18$ & $24^{+5}_{-7}$ & 0 \\ 
        & Gray & $1.22\pm0.04$ & $0.51\pm0.02$ & $2.2_{-0.4}^{+0.5}$ & $3.9_{-0.9}^{+1.1}$ & $59^{+12}_{-10}$ & $1900^{+4400}_{-1000}$ & $1098\pm20$ & $20^{+33}_{-13}$ & -8.4 \\
        & Clear & $0.98_{-0.07}^{+0.04}$ & $0.50_{-0.07}^{+0.03}$ & $3.1_{-0.7}^{+0.9}$ & $6.0_{-1.6}^{+2.1}$ & $54^{+13}_{-11}$ & $1800^{+4700}_{-1000}$ & $1212\pm40$ & $42^{+33}_{-40}$ & -19.4  \\
        \midrule
        HR 8799d & EddySed & $1.09\pm0.03$ & $0.58\pm0.01$ & $9.2_{-2.4}^{+3.3}$ & $2.4_{-1.3}^{+1.9}$ & $108^{+31}_{-22}$ & $1600^{+2700}_{-900}$ & $1202\pm16$ & $17_{-8}^{+9}$ & 0 \\ 
        & Gray & $1.18^{+0.03}_{-0.05}$ & $0.58\pm0.02$ & $4.5^{+2.1}_{-1.6}$ & $1.7^{+1.3}_{-1.0}$ & $107^{+34}_{-24}$ & $1500^{+2800}_{-800}$ & $1161\pm25$ & $19^{+34}_{-14}$ & -2.6 \\
        & Clear & $1.04\pm0.02$ & $0.51\pm0.01$ & $19.3_{-4.5}^{+6.3}$ & $4.4^{+4.0}_{-3.0}$ & $99^{+35}_{-23}$ & $1900^{+4900}_{-1200}$ & $1243\pm15$ & $0.07\pm0.01$ & -13.7 \\
        \midrule
        HR 8799e & EddySed & $0.97^{+0.10}_{-0.13}$ & $0.53^{+0.03}_{-0.04}$ & $1.5_{-0.7}^{+1.5}$ & $1.3_{-1.1}^{+2.0}$ & $30^{+11}_{-8}$ & $>160$ & $1260\pm80$ & $14_{-12}^{+10}$ & 0 \\ 
        & Gray & $0.95\pm0.06$ & $0.52\pm0.04$ & $0.9_{-0.3}^{+0.6}$ & $1.1^{+1.8}_{-1.0}$ & $29^{+12}_{-8}$ & $>140$ & $1276\pm40$ & $12^{+32}_{-9}$ & -4.3 \\
        & Clear & $0.93_{-0.07}^{+0.09}$ & $0.48_{-0.04}^{+0.08}$ & $2.4^{+2.0}_{-1.0}$ & $1.2^{+3.3}_{-1.2}$ & $27_{-8}^{+13}$ & $>125$ & $1295\pm50$ &  $0.6^{+48}_{-0.3}$ & -10.0 \\
        \bottomrule
    \end{tabular}
    } 
     \begin{flushleft}
    {Selected atmospheric parameters and their central 68\% credible interval with equal probability above and below the median are listed. The C/H and S/H values listed are relative to the HR 8799 A abundances from ref. \cite{Baburaj2025}. The values for $\coo$ for planet e are $3\sigma$ lower limits. The C/O ratios reported here account for oxygen sequestration in silicate clouds, following Eq. 12 of ref. \cite{Calamari2024}. For $\ln{B}$, the EddySed cloud model is taken as the baseline model, with the gray and clear models compared against it.}
     \end{flushleft}
\end{table*}

\clearpage

\bibliography{sn-bibliography}

\end{document}